\documentclass[letterpaper,12pt]{article}

\usepackage{graphicx}
\usepackage{amsmath,amsthm,amssymb,mathrsfs,mathtools}
\usepackage[OT1]{fontenc} 
\usepackage{etoc}
\usepackage{float}
\usepackage{fancyhdr}

\def\subsectiontitle{}
\def\subsubsectiontitle{}
\usepackage{tikz}
\usetikzlibrary{automata, positioning}
\usetikzlibrary{arrows,backgrounds}

\tikzstyle{object}=[circle,draw=red]
\tikzstyle{agent}=[circle,draw=blue]
\tikzstyle{quantity}=[fill=white]

\usepackage[noamsmath]{kpfonts}
\usepackage{zi4}

\usepackage{enumitem}
\usepackage[margin=1in]{geometry}

\usepackage[colorlinks,breaklinks=true]{hyperref}

\usepackage{breakcites}

\usepackage{xcolor}
\AtBeginDocument{%
	\hypersetup{%
		linkcolor={blue!50!black},%
		citecolor={blue!50!black},%
		urlcolor={blue!80!black}%
	}%
}%

\usepackage[nameinlink,noabbrev,sort,capitalise]{cleveref}
\makeatletter
\def\ps@pprintTitle{%
	\let\@oddhead\@empty
	\let\@evenhead\@empty
	\def\@oddfoot{\emph{Very preliminary version}\hfill\emph{This draft: \today}}%
	\let\@evenfoot\@oddfoot}
\makeatother
\usepackage{etoolbox}
\patchcmd{\pprintMaketitle}
{\fi\hrule}
{\fi\ifvoid\extrainfobox\else\unvbox\extrainfobox\par\vskip10pt\fi\hrule}
{}{}

\newsavebox\extrainfobox

\usepackage{comment}
\usepackage{multirow}
\usepackage{multicol}
\usepackage{subcaption}
\usepackage{adjustbox}

\allowdisplaybreaks

\let\oldfootnote\footnote
\renewcommand\footnote[1]{\oldfootnote{\hspace{.4mm}#1}}

\makeatletter
\renewenvironment{proof}[1][\proofname] {\par\pushQED{\qed}\normalfont\topsep6\p@\@plus6\p@\relax\trivlist\item[\hskip\labelsep\bfseries#1\@addpunct{.}]\ignorespaces}{\popQED\endtrivlist\@endpefalse}
\makeatother

\let\oldFootnote\footnote
\newcommand\nextToken\relax

\renewcommand\footnote[1]{%
	\oldFootnote{#1}\futurelet\nextToken\isFootnote}

\newcommand\isFootnote{%
	\ifx\footnote\nextToken\textsuperscript{,}\fi}

\usepackage{blkarray}
\usepackage{graphicx,pgfpages,epsfig,multirow,lscape}
\usepackage{hyperref}
\usepackage{epic}
\usepackage{xcolor}
\usepackage{setspace}
\usepackage{tikz}
\usetikzlibrary{arrows,decorations,decorations.pathreplacing,calc,matrix}
\usepackage{natbib}

\DeclareFontFamily{U}{mathb}{\hyphenchar\font45}
\DeclareFontShape{U}{mathb}{m}{n}{
	<-6> mathb5 <6-7> mathb6 <7-8> mathb7
	<8-9> mathb8 <9-10> mathb9
	<10-12> mathb10 <12-> mathb12
}{}
\DeclareSymbolFont{mathb}{U}{mathb}{m}{n}

\newtheorem{definition}{Definition}

\newtheorem{theorem}{Theorem}
\newtheorem*{theorem*}{Theorem}
\newtheorem{proposition}{Proposition}
\newtheorem*{proposition*}{Proposition}
\newtheorem{lemma}{Lemma}
\newtheorem{example}{Example}

\newtheorem{claim}{Claim}

\newtheorem{innercustomexample}{Example}
\newenvironment{customexample}[1]
{\renewcommand\theinnercustomexample{#1}\innercustomexample}
{\endinnercustomexample}

\usepackage{setspace}
\def\w{\omega}

\def\a{\mathcal{A}}

\def\I{\mathcal{I}}

\begin{document}
	
	\title{Rectified strong core in Shapley-Scarf housing markets with indifferences\thanks{I received useful feedback from seminars and conferences at University of Macau, Singapore Management University, Peking University HSBC Business School, Zhejiang University, 2024 Social Choice and Welfare Meeting (Paris). Financial support was provided by the National Natural Science Foundation of China (72122009, 72394391) and the Wu Jiapei Foundation of China Information Economics Society (E21103567).}
    }
	
	\author{Jun Zhang\thanks{Institute for Social and Economic Research, Nanjing Audit University. Email: zhangjun404@gmail.com}
		}
	
	\date{\today}
	
	\maketitle
	
	\begin{abstract}

		We study core concepts in the Shapley-Scarf housing market model under unrestricted weak preferences. Among standard concepts, the strong core may be empty while the nonempty weak core may contain Pareto inefficient allocations. We propose the rectified strong core by adding a single condition: an unaffected agent may join a blocking coalition only if every object he views as indifferent to his current assignment is owned by the coalition. The rectified strong core is always nonempty and Pareto efficient, lies between the weak and strong cores, and coincides with the strong core whenever the latter is nonempty. This condition is the most permissive admissibility requirement for unaffected blocking agents that guarantees nonemptiness, and extends the behavioral foundation of the strong core to weak preferences.
	\end{abstract}

	

	\thispagestyle{empty}
	\setcounter{page}{0}
	\newpage
	
	\section{Introduction}\label{section:Intro}

	Understanding matching models from a cooperative perspective has been a long-standing tradition in market design, dating back to \cite{gale1962college} and \cite{shapley1974cores}. This paper studies the core in the housing market model of \cite{shapley1974cores}, in which unit-demand agents exchange indivisible endowments based on ordinal preferences without money; the model has since become the foundation of a wide range of allocation problems. The work of \cite{shapley1974cores} is well known for introducing the top trading cycles (TTC) mechanism, but their original purpose is to prove the nonemptiness of the core in the presence of indivisibility. The core they prove is later referred to as \textit{weak core}, consisting of allocations that no coalition can strictly improve upon by reallocating members' endowments. The more demanding \textit{strong core} consists of allocations that no coalition can weakly improve upon with at least one member strictly benefiting. The strong core is attractive because it directly rules out Pareto-improving coalitional deviations, including those whose implementation requires the participation of agents whose welfare is unchanged. Under strict preferences, the strong core is nonempty and contains a unique element that can be found via TTC. Under weak preferences, however, the strong core is Pareto efficient yet may be empty, while the weak core is always nonempty yet may contain Pareto inefficient allocations. Pareto efficiency is a key desideratum in many environments, and weak preferences are pervasive in practice and have been considered since the work of \cite{shapley1974cores}. Given the foundational role of the model, this paper revisits these issues and proposes an extension of the strong core that ensures both nonemptiness and Pareto efficiency under weak preferences.

	Weak preferences arise naturally in many important applications. In kidney exchange, it is standard to assume that patients are indifferent among compatible kidneys \citep{roth2005pairwise}. In allocations of multi-unit objects, agents are naturally indifferent among identical copies of the same object. Additional arguments for studying weak preferences can be found in \cite{bogomolnaia2004random}, \cite{bogomolnaia2005strategy}, and \cite{erdil2017two}, among others. The challenge weak preferences pose can be understood by examining the different assumptions under the two core concepts. The weak core adopts the standard assumption that agents are self-interested: an agent joins a blocking coalition only if he strictly benefits. The strong core, however, permits an agent to join a blocking coalition to benefit others as long as his welfare is unaffected. This difference is inconsequential in transferable utility settings, since transfers can make all coalition members strictly better off, but it is crucial in non-transferable settings. In the housing market model, allowing unaffected agents to join blocking coalitions is essential for guaranteeing Pareto efficiency, yet under weak preferences this permissiveness may render the strong core empty. To avoid this, the literature often restricts attention to strict preferences, under which the strong core is nonempty. The strict preference assumption, however, rules out many important applications and obscures the general question of how unaffected agents should be treated in blocking coalitions.

	We develop the following condition for an unaffected agent to join a blocking coalition: he joins only if \textit{all} objects that he views as indifferent to his current assignment are owned by the coalition. Adding this condition to the definition of the strong core, we obtain a new concept referred to as the \textit{rectified strong core}. The added condition is in sharp contrast with the implicit assumption of the strong core that \textit{one} indifferent object suffices to persuade an unaffected agent to join a coalition. 
	The condition automatically holds under strict preferences, since each unaffected agent views his assignment as the  unique indifferent object, which must be owned by the coalition. However, the condition becomes nontrivial under weak preferences, since agents may have expanded indifference classes. From this perspective, the rectified strong core can be viewed as a natural extension of the strong core from strict to weak preferences. Indeed, the rectified strong core is always nonempty, Pareto efficient, a subset of the weak core, and coincides with the strong core whenever the latter is nonempty (Theorem \ref{theorem:rectifiedcore}).

	The coincidence property reflects the normative primacy of the strong core when it is attainable. The strong core is the most demanding pointwise stability concept: it is Pareto efficient and immune to blocking even by coalitions that contain unaffected participants. \cite{quint2004houseswapping} show that the strong core is nonempty only in very special market structures, but in precisely those markets its survival of the broadest class of coalitional blocking makes it the most compelling cooperative solution. It would be counterintuitive to select an outcome that admits weaker deviations when the strong core can be sustained.

	We provide two approaches to justify our proposed condition. The first is a behavioral approach with two complementary arguments. The first argument is a \textit{compellability} argument, inspired by exclusion rights introduced by \cite{balbuzanov2019endowments} for indivisible object allocations. Under strict preferences, every non-redundant unaffected member of a blocking coalition directly or indirectly relies on the endowments of strictly-better-off members to maintain his welfare, so his participation can be interpreted as compelled. Our condition extends this compellability argument to weak preferences for unaffected agents who similarly rely on the endowments of strictly-better-off coalition members to maintain their welfare. Under weak preferences, however, there may exist a group of unaffected coalition members who cannot be compelled because they can maintain their welfare using only their own endowments. For such agents, our condition incorporates a \textit{neutrality} argument: if they were to reclaim their own endowments to maintain their welfare, they would displace exactly the same agents outside the coalition who are displaced by the original blocking. Therefore, their participation produces the same external impact as their withdrawal, while benevolently enabling a Pareto improvement within the coalition.

	The second approach shows that our condition is tight: among all admissibility requirements for unaffected agents satisfying natural regularity conditions, ours is the most permissive one that guarantees nonemptiness of the associated core (Proposition~\ref{prop:participation:tightness}).

  Following the tradition initiated by \cite{shapley1974cores}, we establish the nonemptiness of the rectified strong core constructively. Under weak preferences, TTC after arbitrary tie-breaking produces allocations belonging to the weak core, but these may fail to be Pareto efficient. Several papers have generalized TTC to restore Pareto efficiency and strategy-proofness \citep{alcalde2011exchange,jaramillo2012difference,aziz2012housing,saban2013house,plaxton2013simple,ahmad2021group}. The outcomes of these algorithms belong to the weak core and to the strong core when the strong core is nonempty. We introduce a generalization of TTC that follows the common format of these existing algorithms to restore Pareto efficiency, but imposes no restriction on the pointing rule that needs careful selection to maintain strategy-proofness. We show that all outcomes of the algorithm belong to the rectified strong core (Proposition~\ref{prop:GTTC:rectifiedcore}). In the special setting where agents share common indifferences, a stronger result holds: the original TTC already produces rectified strong core allocations (Proposition~\ref{prop:rectifiedcore:multiplecopies}), establishing a direct connection between this classical algorithm and our new concept.

	We discuss other potential solutions for the housing market model as well. We first examine the \textit{exclusion core} of \cite{balbuzanov2019endowments}, which coincides with the strong core in the housing market model under strict preferences. Under weak preferences, however, it no longer coincides with the strong core and may also be empty; yet we show that it is more often nonempty than the strong core. When nonempty, the exclusion core is always a subset of the rectified strong core (Proposition~\ref{prop:exclusioncore}), establishing a precise relationship between the two concepts.

	We then examine two setwise solutions, the von Neumann-Morgenstern (vNM) stable set of \cite{vonNeumann1944} and the myopic stable set of \cite{demuynck2019myopic}. Whether an allocation belongs to such a set depends on other allocations in the set, unlike pointwise core concepts. In the housing market model, when the strong core is nonempty, it is the unique vNM stable set based on weak domination \citep{wako1991some}. In general, however, the vNM stable set may fail to exist \citep{wako2007nonexistence} or may not be unique. \cite{demuynck2019myopic} show that the myopic stable set coincides with the strong core in the housing market model under strict preferences. We show that under weak preferences, however, the myopic stable set may contain Pareto inefficient allocations.

	Finally, we examine the bargaining set of \cite{yilmaz2022stability}, which is always nonempty and Pareto efficient and lies between the strong and weak cores. However, we show that it may be strictly larger than the strong core when the latter is nonempty, and it is neither a subset nor a superset of the rectified strong core. Unlike core concepts, it is a farsighted concept requiring a two-step verification: for every potential blocking coalition, one must check for the existence of a counter-blocking coalition, which may be computationally more challenging than verifying a core concept.

	Although our analysis is carried out in the housing market model, our approach is not limited to this particular model. The model serves as a canonical setting to isolate a broader issue in cooperative solution concepts for non-transferable-utility environments with weak preferences: when is the participation of unaffected agents in a blocking coalition legitimate? Conventional core concepts give two extreme answers. The weak core effectively excludes unaffected agents from blocking coalitions, which may allow Pareto inefficient allocations to survive. The strong core allows any unaffected agent to participate, which restores efficiency but may destroy existence. This tension is not an artifact of the model's simplicity; it arises in any environment where transfers are limited and weak preferences are present. The housing market model, as the simplest such environment, provides a conceptual benchmark for extensions to richer models. Our proposed condition offers an intermediate principle that links admissible participation to a coalition's control of the relevant resources, and may therefore be useful in other matching and allocation problems where agents have weak preferences and coalitional deviations depend on the resources, positions, or rights controlled by the deviating coalition.

	The remainder of the paper is organized as follows. Section~\ref{section:preliminaries} defines the model and the standard core concepts. Section~\ref{section:rectified:core} introduces the rectified strong core and justifies our proposed condition. Section~\ref{section:GTTC} presents the algorithm to produce rectified strong core allocations. Section~\ref{section:othersolutions} examines alternative cooperative solution concepts. Section \ref{section:conclusion} concludes the paper. The appendix includes all proofs and additional results.

	\paragraph{Related literature}

	The housing market model has been extensively studied; see \cite{afacan2024housing} for a survey. Although \cite{shapley1974cores} and \cite{roth1977weak}, the two seminal papers on the model, both consider unrestricted preferences, the subsequent literature has mostly focused on strict preferences. Among the papers addressing weak preferences, some generalize TTC to preserve Pareto efficiency and strategy-proofness, and others study competitive allocations, those produced by TTC after arbitrary tie-breaking. However, competitive allocations exhibit the anomalies emphasized by \cite{roth1977weak}: the set of competitive allocations may be strictly larger than a nonempty strong core; in some markets all competitive allocations are Pareto inefficient; there may exist two allocations that are indifferent to all agents, but one is competitive yet the other is not. We discuss these and other related studies below.

	\cite{wako1984note} shows that in the housing market model, a nonempty strong core may be a strict subset of the set of competitive allocations, and \cite{wako1991some} shows that the two coincide if and only if any two competitive allocations are indifferent to all agents. \cite{quint2004houseswapping} provide a necessary and sufficient condition for the strong core to be nonempty; we utilize this result when studying the relationship between the rectified strong core and the strong core. \cite{biro2023shapley} prove that the strong core satisfies the respect-improvement property whenever it is nonempty,\footnote{The respect-improvement property requires that an agent receive a weakly better assignment when his endowment becomes more desirable to others.}  and that the set of competitive allocations satisfies that property in a stochastic-dominance sense.

	Several papers provide cooperative foundations for the set of competitive allocations. \cite{wako1999coalition} shows that it coincides with a modification of the strong core based on antisymmetric weak domination,\footnote{It requires that every unaffected agent in a blocking coalition must retain his original assignment.} and \cite{toda1997implementation} shows that it is the unique vNM stable set under the same domination; \cite{kawasaki2010farsighted} and \cite{klaus2010farsighted} extend this to farsighted versions, though these results require additional assumptions on preferences or domination, because a vNM stable set generally does not exist in the housing market model \citep{wako2007nonexistence}. \cite{ehlers2004monotonic} shows that the set of competitive allocations is the minimal monotonic extension of the strong core.\footnote{A monotonic extension of the strong core is a superset of the strong core that satisfies Maskin's Monotonicity, which is necessary for Nash implementation.} 
	 
	 Several papers generalize TTC to weak preferences. \cite{alcalde2011exchange} and \cite{jaramillo2012difference} independently propose the first such generalizations; \cite{aziz2012housing} and \cite{plaxton2013simple} study their computational properties. \cite{saban2013house} unify these algorithms into a family parameterized by pointing rules and derive sufficient conditions for strategy-proofness; \cite{ahmad2021group} study weak group strategy-proofness. Our algorithm follows the common format of these algorithms but imposes no restriction on pointing rules. If one insists on strategy-proofness, the existing pointing rules can be imposed. 
	 
	 \cite{sandholtz2025shapley} study the special setting we consider in Section \ref{section:GTTC}, in which all agents share common indifferences. They show that TTC with fixed tie-breaking is Pareto efficient and core-selecting, a result implied by our Proposition~\ref{prop:rectifiedcore:multiplecopies}.

  \cite{balbuzanov2019endowments} introduce the exclusion core in a general indivisible object allocation model and show that it coincides with the strong core in the housing market model under strict preferences. Several papers extend the exclusion core to other settings and still assume strict preferences, including production economies, priority-based allocation, and refinements satisfying consistency \citep{balbuzanov2019property,cheng2024proper,ishida2025group,zhangrefinedexclusioncore}. This paper is the first to examine the exclusion core under weak preferences.

	 Finally, a separate literature studies core concepts in the housing market model with externalities under strict preferences \citep{mumcu2007core,graziano2020shapley,hong2022core,dougan2011core,aslan2020competitive,klaus2023core}. This literature shows that even the weak core may be empty under externalities, so additional preference restrictions are needed to restore nonemptiness.

  \section{Preliminaries}\label{section:preliminaries}
	
	\subsection{The Shapley-Scarf housing market model}\label{section:model}

	In the housing market model, a market is represented by a tuple $M= (I,O,\w,\succsim_I) $, where $ I $ is a finite set of agents, $ O $ is a finite set of objects with $ |O|=|I| $, $ \w $ is a one-to-one mapping from $ I $ to $ O $, and $ \succsim_I=(\succsim_i)_{i\in I} $ is a preference profile. For each $ i\in I $, $ \w(i) $ is the object owned by $ i $. Each agent $ i $ has a preference relation $ \succsim_i $ over $ O $ that is transitive and complete but not necessarily antisymmetric. 	
	For any distinct $ o ,o'\in O $, we write $ o\succ_i o' $ if $ o\succsim_i o' $ but $ o'\not\succsim_i o $, meaning that $ i $ strictly prefers $ o $ over $ o' $; we write $ o\sim_i o' $ if $ o\succsim_i o' $ and $ o'\succsim_i o $,  meaning that $ i $ is indifferent between $ o $ and $ o' $.  Every nonempty  $ C\subseteq I$ is called a \textbf{coalition}. A coalition $ C' $ is a \textbf{subcoalition} of $ C $ if $ C'\subseteq C $. 	
	For convenience, we often write a market as $ M(\w,\succsim_I) $ to highlight its initial endowment distribution and preferences.

	An \textbf{allocation} is a one-to-one mapping $\mu: I \rightarrow O $, where $\mu(i)$ denotes the object received by $i$. An allocation $ \mu $ is \textbf{individually rational} if,  $\forall i\in I $, $ \mu(i) \succsim_i \w(i) $. 
	Two allocations $ \mu $ and $ \mu' $ are \textbf{indifferent} if, $\forall i\in I $, $ \mu(i) \sim_i \mu'(i) $.  An allocation $ \sigma $ is a \textbf{Pareto improvement} over another $ \mu $ for a coalition $ C $ if, $\forall i\in C $, $ \sigma(i)\succsim_i \mu(i) $, and for some $ j\in C $, $ \sigma(j)\succ_j \mu(j) $. An allocation is \textbf{Pareto efficient} if it admits no Pareto improvement for the grand coalition $I$. 
	
	We introduce the following notation. For each $ i\in I $ and each $ O'\subseteq O $, let $ B_i(O')=\{o\in O':o\succsim_i o' \text{ for all }o'\in O'\} $ denote the set of $ i $'s most preferred objects in $ O' $. For each $ o\in O $, let $ \mathcal{I}_i(o)=\{o'\in O: o'\sim_i o\}$ denote the set of objects that $ i $ views as indifferent to $ o $. So, $\I_i(o)$ is the indifference class in $i$'s preferences that includes $o$. For each coalition $ C $ and each allocation $ \mu $, let $ \mu(C)=\cup_{i\in C}\{\mu(i)\}$ denote the set of objects assigned to $ C $. By viewing $ \w $ as an initial allocation, $ \w(C) $ is the set of objects initially owned by $ C $.
	
	Let $ \mathcal{M} $ denote the set of markets. For every $ M\in \mathcal{M} $, let $ \mathcal{A}(M) $ denote the set of allocations and $ 2^{\mathcal{A}(M)} $ its power set. A \textbf{solution} is a correspondence $ f:\mathcal{M}\rightarrow \bigcup_{M\in \mathcal{M}}2^{\mathcal{A}(M)}$ such that, for every $ M\in \mathcal{M} $, $ f(M)\in 2^{\mathcal{A}(M)} $; we allow $ f(M) $ to be empty for some $ M $. A solution $ f $ is \textit{Pareto efficient} if, for every $ M\in \mathcal{M} $, whenever $ f(M) $ is nonempty, all elements of $ f(M) $ are Pareto efficient allocations in $M$.

\cite{shapley1974cores} introduce the TTC mechanism for the housing market model. Under strict preferences, TTC proceeds as follows: in each step, each remaining agent points to his most preferred object and each remaining object points to its owner; the agents in each resulting cycle exchange their endowments and then leave; the procedure stops when all agents have left. Under strict preferences, TTC produces a unique allocation. Under weak preferences, ties must be broken before running the algorithm, and different tie-breaking rules may yield different allocations. We refer to all such allocations as TTC outcomes under weak preferences.

	\subsection{Strong core and weak core}\label{section:standard:core}
	
	In the standard definition of the core, a coalition blocks an allocation by reallocating its members' endowments among themselves. Depending on whether all coalition members must strictly benefit from the blocking, the definition has two variants.

	\begin{definition}\label{definition:strongcore}
		In a market $ M(\w,\succsim_I) $, an allocation $ \mu $ is \textbf{weakly blocked} by a coalition $ C $ via another allocation $ \sigma $ if
		\begin{enumerate}
			\item $ \forall i\in C $, $ \sigma(i) \succsim_i \mu(i) $ and $ \exists j\in C $, $ \sigma(j)\succ_j\mu(j) $;
			
			\item $ \sigma(C)= \w(C)$.
		\end{enumerate}
	The \textbf{strong core} consists of all allocations that are not weakly blocked.
	
	If condition (1) is replaced by ``$ \forall i\in C $, $ \sigma(i) \succ_i \mu(i) $,'' then we say that $ \mu $ is \textbf{strongly blocked} by $ C $ via $ \sigma $. The \textbf{weak core} consists of all allocations that are not strongly blocked.
	\end{definition}

	For any coalition $C$ that (weakly or strongly) blocks an allocation $\mu$ via another $\sigma$, we define $C_{\sigma>\mu}=\{i\in C: \sigma(i)\succ_i \mu(i)\}$ and $C_{\sigma\sim \mu}=\{i\in C: \sigma(i)\sim_i \mu(i)\}$. Here, $C_{\sigma>\mu}$ is the set of \textbf{better-off members} of $C$ and $C_{\sigma\sim \mu}$ is the set of \textbf{unaffected members} of $C$. Strong blocking excludes unaffected agents from a blocking coalition; that is, $C=C_{\sigma>\mu}$. Weak blocking allows $C_{\sigma\sim \mu}$ to be nonempty, as long as $C_{\sigma>\mu}$ is nonempty. Consequently, the weak core is a superset of the strong core.
	
	The weak core is nonempty regardless of whether preferences are strict or weak, but it may contain Pareto inefficient allocations: because unaffected agents are excluded from coalitions, some Pareto improvements cannot be enforced. The strong core guarantees Pareto efficiency whenever it is nonempty. Under strict preferences, the strong core is nonempty and consists of a unique allocation that is found via TTC. Under weak preferences, however, the strong core may be empty, because weak blocking sets an overly permissive condition for unaffected agents to join a blocking coalition, that is, as long as they receive an object indifferent to their original assignments. Under weak preferences, indifference can be widespread, so this condition can be easily satisfied. An agent with many indifferent objects can join numerous blocking coalitions, obstructing the existence of strong core allocations. Example \ref{example:strongcore:empty} illustrates this.

	\begin{example}[Strong core may be empty whereas weak core is overly large]\label{example:strongcore:empty}
		Three agents $ 1,2,3$ respectively own three objects $ a,b,c $. The left table below lists agents' endowments and the allocations under consideration. The other two tables list two preference profiles.
		\begin{table}[!h]
			\centering
			\begin{subtable}[b]{.3\linewidth}
				\centering
				\begin{tabular}{cccc}
					& $ 1 $ & $ 2 $ & $ 3 $  \\ \hline
					$ \w $: & $ a $ & $ b $ & $ c $ \\ \hline
					$ \mu $: & $ b $ & $ a $ & $ c $ \\
					$ \sigma $: & $ c $ & $ a $ & $ b $ \\
					$ \delta $: & $ a $ & $ c $ & $ b $
				\end{tabular}
			\subcaption{Allocations}
			\end{subtable}
			\begin{subtable}[b]{.3\linewidth}
				\centering
				\begin{tabular}{ccc}
					$ \succsim_1 $	& $ \succsim_2 $ & $ \succsim_3 $  \\ \hline
					$ c $ & $ a $ & $ b $ \\
					$b $ & $ b $ & $ c $ \\
					$ a $ & $ c $ & $ a $\\
					& \\
				\end{tabular}
			\subcaption{$ \succsim_I $}
			\end{subtable}
			\begin{subtable}[b]{.3\linewidth}
				\centering
				\begin{tabular}{ccc}
					$ \succsim'_1 $	& $ \succsim'_2 $ & $ \succsim_3 $  \\ \hline
					$ b $ & $ a,b,c $ & $ b $ \\
					$ a $ &  & $ c $ \\
					$ c $ & & $ a $\\
					& \\
				\end{tabular}
			\subcaption{$ \succsim'_I $}
			\end{subtable}
		\end{table}

		\noindent (\textbf{Strict preferences}) Under $ \succsim_I $, agents have strict preferences and their most preferred objects are distinct, so $ \sigma $ is the unique Pareto efficient allocation. 
		
		The weak core equals $ \{\mu,\sigma\} $, whereas the strong core equals $ \{\sigma\} $. $\mu$ is weakly blocked by $\{1,2,3\}$ via $\sigma$, but not strongly blocked: because agent 2 already holds his favorite object $a$, no strong blocking coalition can include 2 to reallocate $b$, and the coalition $\{1,3\}$ with endowments $\{a,c\}$ cannot make both members strictly better off.
	
		\medskip
		
		\noindent (\textbf{Weak preferences}) Under $ \succsim'_I $, both $ 1 $ and $ 3 $ most prefer $ 2 $'s endowment and then their own, while $ 2 $ is indifferent among all objects. Both $ \mu $ and $ \delta $ are Pareto efficient, yet $\w$ is not.

		The weak core equals $ \{\w, \mu, \delta\} $. 		
		To block the Pareto inefficient allocation $\w$, we need a coalition that contains $2$ to reallocate $b$. Because $ 2 $ is indifferent among all objects, however, he cannot join any strong blocking coalition. So, $\w$ is not strongly blocked.

		Nevertheless, the strong core is empty. Allocation $\w$ is weakly blocked by $\{1,2,3\}$ via either $\mu$ or $\delta$. However, $\mu$ and $\delta$ are also weakly blocked: $ \mu $ is weakly blocked by $ \{2,3\} $ via $ \delta $, and $ \delta $ is weakly blocked by $ \{1,2\} $ via $ \mu $. Here, because $ 2 $ is indifferent between $ a $ and $ c $, he alternates between $ 1 $ and $ 3 $ in forming weak blocking coalitions; agent $2$ joins too many blocking coalitions, rendering the strong core empty.
	\end{example}

	\cite{quint2004houseswapping} provide a necessary and sufficient condition for the strong core to be nonempty; see Lemma \ref{lemma:strong:core:nonempty} in Appendix \ref{appendix:Theorem1}. The condition implies that the strong core is nonempty only for a very special market structure, and a nonempty strong core may not be a singleton but must be essentially single-valued: all allocations within it are mutually indifferent and every allocation indifferent to them belongs to the strong core.

\section{Rectified strong core}\label{section:rectified:core}

\subsection{Definition and main result}

Given the limitations of the standard core concepts, our goal is to find a new core concept that guarantees both nonemptiness and Pareto efficiency, and, given the attractiveness of the strong core, we want it to coincide with the strong core whenever the latter is nonempty. As the previous section shows, allowing unaffected agents to participate in blocking coalitions is necessary for Pareto efficiency, yet their participation must be restricted under weak preferences to ensure nonemptiness. The central question therefore becomes when an unaffected agent may legitimately join a blocking coalition. We propose the following \textit{no outside option} condition: an unaffected agent may join a blocking coalition only if \textit{all} objects that he views as indifferent to his current assignment are owned by the coalition. Our proposed core concept is then defined as follows.

\begin{definition}\label{definition:rectified:core}
	In a market $ M(\w,\succsim_I) $, an allocation $ \mu $ is \textbf{rectification blocked} by a coalition $ C $ via another allocation $ \sigma $ if
	\begin{enumerate}
		\item $ \forall i\in C $, $ \sigma(i) \succsim_i \mu(i) $, and $ \exists j\in C $, $ \sigma(j)\succ_j\mu(j) $;
		
		\item $ \sigma(C)= \w(C)$;
		
		\item $ \forall i\in C_{\sigma\sim\mu} $, $ \mathcal{I}_i(\mu(i))\subseteq \w(C)$.
	\end{enumerate}
	The \textbf{rectified strong core} consists of allocations that are not rectification blocked.
\end{definition}

Condition~(3) formalizes the no outside option condition. It is the only addition relative to the definition of the strong core. The strong core implicitly assumes that \textit{one} indifferent object suffices to persuade an unaffected agent to join a blocking coalition, whereas condition~(3) tightens this by requiring the coalition to own \textit{all} of his indifferent objects. Under strict preferences, every unaffected agent $i\in C_{\sigma\sim\mu}$ satisfies $\sigma(i) = \mu(i)$. Since $\sigma(C) = \omega(C)$, we have $\mu(i) \in \omega(C)$, and therefore $\mathcal{I}_i(\mu(i)) = \{\mu(i)\} \subseteq \omega(C)$. So, condition (3) holds automatically under strict preferences, and the two cores coincide. 
Under weak preferences, agents' indifference classes may expand: $\mathcal{I}_i(\mu(i))$ may contain objects outside $\omega(C)$. The rectified strong core is thus the definition obtained by making condition (3) explicit. From this perspective, the rectified strong core is a natural extension of the strong core from strict to weak preferences.

By definition, the rectified strong core is a superset of the strong core. When the strong core is empty, the rectified strong core is a strict superset of the strong core. For instance, in Example \ref{example:strongcore:empty}, under $\succsim'_I$, the strong core is empty, whereas the rectified strong core equals $\{\mu,\delta\}$: $\mu$ and $\delta$ cannot be rectification blocked via each other, because agent $2$ always has an indifferent object not owned by the blocking coalition considered before. When the strong core is nonempty, we show that the rectified strong core does not recommend any allocations outside the strong core; they coincide. So, the following theorem summarizes the properties of the rectified strong core.

\begin{theorem}\label{theorem:rectifiedcore}
	The rectified strong core is nonempty and Pareto efficient. It is a subset of the weak core and coincides with the strong core whenever the latter is nonempty.
\end{theorem}

We prove nonemptiness constructively: the allocations produced by the algorithm in Section \ref{section:GTTC} belong to the rectified strong core. The proof for the coincidence with a nonempty strong core employs the condition of \cite{quint2004houseswapping}; see Appendix \ref{appendix:Theorem1}.

 In the next two subsections, we provide two approaches to justify condition~(3)  of Definition \ref{definition:rectified:core}. The first is a behavioral approach.  
 The second approach shows that condition~(3) is tight: the condition may appear stringent, but any weaker admissibility requirement for unaffected agents may fail to guarantee nonemptiness of the associated core concept.

\subsection{Behavioral approach to justify condition (3)  of Definition \ref{definition:rectified:core}}

\paragraph{Strict preferences.} 
Under strict preferences, condition (3) imposes no additional restriction beyond conditions (1) and (2), yet it is instructive to examine why unaffected agents' participation is justified, as the argument extends naturally to weak preferences.

Suppose that a coalition $ C $ weakly blocks an allocation $ \mu $ via another $ \sigma $, and $ C_{\sigma\sim\mu}\neq \emptyset $. If there exists a subcoalition $C'\subseteq C_{\sigma\sim\mu}$ such that $\sigma(C')=\w(C')$, then we call $C'$ \textit{redundant}, since its members allocate their endowments among themselves under both $\mu$ and $\sigma$, and therefore removing them has no impact on remaining agents within the coalition. For every remaining unaffected agent $ i_1\in C_{\sigma\sim\mu} $, since $\sigma(i_1)=\mu(i_1)$ and $\sigma(C)=\w(C)$, we must have $\mu(i_1)=\w(i_2)$ for some $i_2\in C$. If $i_2\in C_{\sigma>\mu}$, it means that $i_1$ receives the endowment of a strictly-better-off agent in $C$. If $i_2\in C_{\sigma\sim\mu}$, then the same reasoning applies and we must obtain a chain:
\[
i_1 \rightarrow i_2 \rightarrow \cdots \rightarrow i_K\rightarrow j,
\]
in which $i_k\in C_{\sigma\sim\mu}$ and $\mu(i_k)=\w(i_{k+1})$ for every $k$, and $\mu(i_K)=\w(j)$ for some $j\in C_{\sigma>\mu}$. Since $C$ is finite, the chain must terminate at a strictly-better-off agent in $C$. The existence of such a chain means that every $i_1\in C_{\sigma\sim\mu}$ directly or indirectly relies on the endowments of $C_{\sigma>\mu}$ to remain unaffected, which justifies their participation in the coalition.

Specifically, if $j$ were to reclaim his endowment, then $i_K$ would be unable to remain unaffected. So, $i_K$ relies on $j$'s endowment to maintain his welfare and therefore can be viewed as compelled by $j$ to join the coalition. Similarly, once $i_K$ joins the coalition, if $i_K$ were to reclaim his endowment, $i_{K-1}$ would be unable to remain unaffected. So, $i_{K-1}$ directly relies on $i_K$'s endowment and indirectly relies on $j$'s endowment to maintain his welfare, and therefore he can be viewed as compelled to join the coalition. This \textit{compulsion} propagates back along the chain, until $i_1$ is compelled as well. In the terminology of \cite{balbuzanov2019endowments}, which is discussed in Section \ref{section:exclusioncore}, the endowments of all agents in the chain, including $i_1$'s,  are controlled by $j$ and under $j$'s compulsion.

\paragraph{Weak preferences.} Under weak preferences, two complications arise. First, an agent's indifference class $\mathcal{I}_i(\mu(i))$ may contain multiple objects. Conditions~(1) and~(2) in Definition~\ref{definition:rectified:core} guarantee only that, for every $i\in C_{\sigma\sim\mu}$, $\sigma(i)\in\mathcal{I}_i(\mu(i))\cap\w(C)\neq\emptyset$; they do not prevent other objects in $\mathcal{I}_i(\mu(i))$ from lying outside $\w(C)$. If such an object $o\in\mathcal{I}_i(\mu(i))\backslash\w(C)$ exists, then $i$ could remain unaffected by consuming $o$ without relying on the coalition's endowments. Second, and more seriously, an agent may be able to remain unaffected by reclaiming his own endowment, and thus cannot be compelled. More generally, a subset of unaffected agents may be able to allocate their endowments among themselves in a way that maintains their welfare, making them self-sufficient without relying on others.

Condition~(3) addresses both complications in a unified way. By requiring $\mathcal{I}_i(\mu(i))\subseteq\w(C)$ for every $i\in C_{\sigma\sim\mu}$, condition (3) is equivalent to requiring
\[
\forall i\in C_{\sigma\sim\mu}, \quad \mathcal{I}_i(\mu(i))\backslash \w(C_{\sigma\sim\mu})\subseteq\w(C_{\sigma>\mu}).
\] 
That is, every welfare-preserving object for every $i\in C_{\sigma\sim\mu}$ that is not owned by $C_{\sigma\sim\mu}$ is owned by $C_{\sigma>\mu}$. Depending on agents' preferences, $C_{\sigma>\mu}$ can manifest its control over the welfare of $C_{\sigma\sim\mu}$ in two ways. 

When no subcoalition of $C_{\sigma\sim\mu}$, including $C_{\sigma\sim\mu}$ itself, can maintain its members' welfare using only its own endowments, $C_{\sigma>\mu}$'s control translates into compellability: $C_{\sigma>\mu}$ can threaten to withdraw their endowments, forcing every unaffected agent to lose welfare. This extends the \textit{compellability} argument under strict preferences to allow for indifferences. Example \ref{example:extremecase:1} illustrates this case.

In another extreme case, $C_{\sigma\sim\mu}$ itself can reallocate its own endowments to be self-sufficient, so the compellability argument breaks down. But condition (3) implies that if $C_{\sigma>\mu}$ reclaims its endowments, $C_{\sigma\sim\mu}$ would be forced to reclaim $\w(C_{\sigma\sim\mu})$, displacing exactly the outside agents who lose their assignments under the original blocking.\footnote{In the original blocking, conditions (1) and (2) of Definition~\ref{definition:rectified:core} do not refer to the assignments for the agents outside $C$. We can always choose a blocking allocation $\sigma$ such that, for every $j\in I\backslash C$, $\sigma(j)\neq \mu(j)$ if and only if $\mu(j)\in \w(C)$. That is, the agents outside $C$ who are affected by the blocking are only those who receive the coalition's endowments under $\mu$.} The participation of $C_{\sigma\sim\mu}$ is therefore \textit{neutral}: it does not alter the external impact on outside agents, while enabling a Pareto improvement within $C$. Example \ref{example:extremecase:2}	illustrates this case.

In general, agents' preferences may fall between the above two cases: some subcoalition of $C_{\sigma\sim\mu}$ can maintain its welfare using only its own endowments, while the remaining members of $C_{\sigma\sim\mu}$ must rely on the endowments of $C_{\sigma>\mu}$ to remain unaffected. The participation of the latter group can then be interpreted as compelled, while the participation of the former group can be justified by the neutrality argument. Example \ref{example:othercases} in Appendix \ref{appendix:examples} illustrates this intermediate case.

\begin{example}[Compellability argument]\label{example:extremecase:1}
		Consider the following market with six agents.
		
		\begin{table}[!h]
    \raggedright
    \begin{subtable}[t]{.35\linewidth}
        \begin{tabular}[t]{ccccccc}
            & $ 1 $ & $ 2 $ & $ 3 $ & $ 4 $ & $ 5 $ & $6$ \\ \hline
            $ \w $: & $ a $ & $ b $ & $ c $ & $ d $ & $ e $ & $f$\\ \hline
            $ \mu $: & $ e $ & $ f $ & $ a $ & $ b $ & $ c $ & $d$\\
            $ \sigma $: & $ c $ & $ d $ & $ a $ & $ b $ & $ e $ & $f$\\
        \end{tabular}
    \end{subtable}%
    \begin{subtable}[t]{.35\linewidth}
        \begin{tabular}[t]{cccccc}
            $ \succsim_1 $ & $ \succsim_2 $ & $ \succsim_3 $ & $ \succsim_4 $ & $ \succsim_5 $ & $ \succsim_6 $ \\ \hline
            $ c $ & $ d $ & $ a,b $ & $ b,c $ & $ c,d $ & $c,d$\\
            $ e $ & $ f $ & $ c $ & $ d $ & $ a,b $ & $ a,b$ \\
            $ \vdots $ & $ \vdots $ & $ \vdots $ & $ \vdots $ & $ \vdots $ & $ \vdots $
        \end{tabular}
    \end{subtable}%
    \begin{subtable}[t]{.3\linewidth}
        \centering
				\small
        \begin{tikzpicture}[semithick,bend angle=30,xscale=0.7,yscale=0.6,baseline=(current bounding box.north)]
            \node at (5,-2.8) {$\mu$};
            \node (i1) at (7,0) [agent] {$1$};
            \node (i2) at (7,-1.5) [agent] {$2$};
            \node (i3) at (5,0) [agent] {$3$};
            \node (i4) at (5,-1.5) [agent] {$4$};
            \node (i5) at (3,0) [agent] {$5$};
						\node (i6) at (3,-1.5) [agent] {$6$};
            \draw[-latex] (i1) [bend right] to (i5);
            \draw[-latex] (i2)  [bend left] to (i6);
            \draw[-latex] (i3) to (i1);
            \draw[-latex] (i4) to (i2);
            \draw[-latex] (i5) to (i3);
						\draw[-latex] (i6) to (i4);
        \end{tikzpicture}
    \end{subtable}
\end{table}

Both the strong core and the rectified strong core equal $\{\sigma\}$. Under $\sigma$, each of the four agents in $\{ 1,2,3,4\} $  receives a most preferred object.

The allocation $\mu$ is rectification blocked by $C=\{ 1,2,3,4\} $ via $\sigma$, with $C_{\sigma\sim\mu}=\{3,4\}$ and $C_{\sigma>\mu}=\{1,2\}$. Although $3$ receives $1$'s endowment under $\mu$, $3$ does not rely only on $1$'s endowment to maintain welfare, since he is indifferent between $a$ and $b$. Similarly, $4$ does not rely only on $2$'s endowment to maintain welfare, since he is indifferent between $b$ and $c$. However, the two agents cannot rely only on their own endowments to both maintain welfare; they must rely on the endowments of $\{1,2\}$. So, condition (3) allows them to join the coalition $C$.
\end{example}

\begin{example}[Neutrality argument]\label{example:extremecase:2}		
		Consider the following market with three agents.
		
		\begin{table}[!h]
			\centering
			\begin{subtable}[b]{.25\linewidth}
				\begin{tabular}[b]{cccc}
					& $ 1 $ & $ 2 $ & $ 3 $   \\ \hline
					$ \w $: & $ a $ & $ b $ & $ c $  \\ \hline
					$ \mu $: & $ c $ & $ a $ & $ b $  \\
					$ \sigma $: & $ b $ & $ a $ & $ c $ \\
				\end{tabular}
			\end{subtable}
			\begin{subtable}[b]{.25\linewidth}
				\begin{tabular}[b]{ccc}
					$ \succsim_1 $	& $ \succsim_2 $ & $ \succsim_3 $   \\ \hline
					$ b $ & $ a,b $ & $ b $  \\
					$ c $ & $ c $ & $ c$  \\
					$ a $ &  & $ a $  \\
				\end{tabular}
			\end{subtable}
			\begin{subtable}[b]{.3\linewidth}
				\begin{tikzpicture}[semithick,bend angle=30,xscale=0.8,yscale=0.8]
					\node    at (5,-1.2)  {$\mu$};
					\node (i1) at (3,0) [agent] {$1$};
					\node (i2) at (5,0) [agent] {$2$};
					\node (i3) at (7,0) [agent] {$3$};
					
					\draw[-latex] (i1) [bend right] to (i3);
					\draw[-latex] (i2)  to (i1);
					\draw[-latex] (i3)  to (i2);
				\end{tikzpicture}
			\end{subtable}
		\end{table}
		
		Both the strong core and the rectified strong core equal $\{\sigma\}$. Under $\sigma$, both agent $1$ and agent $2$ receive a most preferred object.

		The allocation $\mu$ is rectification blocked by $C=\{1,2\}$ via $\sigma$, with $C_{\sigma\sim\mu}=\{2\}$ and $C_{\sigma>\mu}=\{1\}$. Agent $2$ cannot be compelled by agent $1$, since he is indifferent between $a$ and his endowment $b$. However, if $1$ reclaims his endowment $a$ from $2$, $2$ would be forced to reclaim $b$, which forces $3$ to forgo $b$. Therefore, although $1$ cannot compel $2$, $1$ can force $\{1,2\}$ to allocate their endowments among themselves. In this case, condition (3) permits $2$ to join the coalition with $1$ to reallocate their endowments in a Pareto efficient way.
	\end{example}

\subsection{Tightness of condition (3) of Definition \ref{definition:rectified:core}}

Condition~(3) requires a blocking coalition to own every object in every unaffected member's indifference class. This is a strong requirement. One may ask whether a weaker requirement still guarantees nonemptiness. That is, retaining conditions~(1) and~(2), we ask how permissive the admissibility requirement for unaffected members can be while preserving nonemptiness of the core concept. The following formulation shows that among a reasonable class of requirements, condition (3) is already the most permissive one.

We employ a function to formulate the admissibility requirement for unaffected coalition members. Given  $I$ and $O$ in a market, a \textbf{participation function} is a mapping
\[
p:\{(X,Z): \emptyset\neq Z\subseteq X\subseteq O\}\rightarrow \{0,1\}.
\]
Here $ X $ is interpreted as the set of objects an unaffected agent views as indifferent to his current assignment, and $ Z $ is the set of objects within the indifference class that are owned by a coalition $C$. If $ p(X,Z)=1 $, it means that an unaffected agent with the indifference class $ X $ is permitted to join $C$ if $C$ owns exactly the objects in $ Z $ from that class; if $ p(X,Z)=0 $, it means that the unaffected agent is not permitted to join $C$.  

We impose two regularity conditions on $p$: \textit{anonymity} requires that if $ |X|=|X'| $ and $ |Z|=|Z'| $, then $ p(X,Z)=p(X',Z') $; \textit{full-coverage sufficiency} requires that, for every $X\subseteq O$, $p(X,X)=1$. Anonymity means that the function depends only on the numbers of indifferent objects in $X$ and coalition-owned indifferent objects in $Z$, not on the labels of objects. Full-coverage sufficiency means that if the coalition owns every welfare-preserving object of an unaffected agent, participation is always permitted. Under strict preferences, this condition imposes no restriction beyond conditions (1) and (2), so that the associated core concept degenerates to the strong core. We call $p$ \textit{regular} if it satisfies the two conditions.

Given a regular participation function $ p $, we define \textbf{$ p $-blocking} by replacing condition~(3) of Definition~\ref{definition:rectified:core} with the requirement that
\[
\forall i\in C_{\sigma\sim\mu}, \quad p\big(\mathcal{I}_i(\mu(i)),\,\mathcal{I}_i(\mu(i))\cap \w(C)\big)=1.
\]
The associated \textbf{$ p $-core} consists of allocations that are not $ p $-blocked. The rectified strong core is associated with the function $p$ such that $p(X,Z)=1$ if and only if $Z=X$.

The next proposition formalizes our observation that condition~(3) is not merely intuitive but most permissive for nonemptiness of the $p$-core.

\begin{proposition}\label{prop:participation:tightness}
	Let $ p $ be any regular participation function. The $p$-core is always nonempty if and only if it is the rectified strong core.
\end{proposition}

In the proof in Appendix \ref{appendix:proof:tightness}, we show that if there exist a nonempty set $ X\subseteq O  $ and a nonempty strict subset $ Z\subsetneq X$ such that $ p(X,Z)=1 $, then there exists a housing market in which the $ p $-core is empty. Therefore, to ensure nonemptiness, we must have $p(X,Z)=1$ if and only if $Z=X$, which is precisely the function defining the rectified strong core.

\section{Finding rectified strong core allocations}\label{section:GTTC}

Under strict preferences, TTC is a Pareto efficient and strategy-proof algorithm, and its unique outcome belongs to the strong core. Under weak preferences, however, its outcomes may not be Pareto efficient. In fact, there exist markets in which all outcomes of TTC are Pareto inefficient. Therefore, those outcomes do not belong to the rectified strong core. This section presents a \textbf{generalized top trading cycles} (GTTC) algorithm. We prove that every allocation it produces belongs to the rectified strong core.
	
 The literature has proposed several generalizations of TTC to weak preferences that preserve Pareto efficiency and strategy-proofness. These algorithms share a common format. Specifically, each step of these algorithms consists of three stages. In the \textit{departure} stage, a group of agents is removed with their current assignments if there are no beneficial trades involving them in subsequent steps. In the \textit{pointing} stage, a pointing rule assigns each agent a unique pointee such that at least one beneficial trading cycle is formed. The choice of the pointing rule is crucial for preserving strategy-proofness. In the \textit{trading} stage, the cycles formed in the pointing stage are traded. 
	
	Our definition of GTTC follows this format but does not specify a pointing rule for the pointing stage. It requires only that at least one beneficial trading cycle be formed. We prove that this requirement suffices to ensure that every outcome of the algorithm belongs to the rectified strong core. If one wants to achieve strategy-proofness, pointing rules from the existing algorithms can be applied.

	\begin{center}
		\textbf{Generalized Top Trading Cycles}
	\end{center}

	\textbf{Step} $ t\ge 1 $: Every step includes three stages. 
	\begin{itemize}
		\item \textbf{Departure}: Among the remaining agents, a group is chosen to depart with the objects they hold if, for every member of the group, two conditions are met:
		
		\begin{enumerate}
			\item He holds one of his most preferred objects among the remaining ones;
			
			\item All of his most preferred objects among the remaining ones are held by the group.
		\end{enumerate}
		
		Once a group departs, there may exist another group that satisfies the same two conditions. Choose one such group and let it depart. Repeat this operation until no further groups can be chosen. 
		At any point in this process, an agent is said to become \textit{satisfied} if the object he holds becomes one of his most preferred objects among the remaining ones.
		
		At the end of the departure stage, denote the set of remaining agents by $ I_t $ and the set of remaining objects by $ O_t $. If $ I_t $ is empty, stop the algorithm. Otherwise, denote the current allocation by $ \mu_t $ such that, for every $ i \in I_t$, $ \mu_t(i) $ is the object held by $ i $. An agent $ i\in I_t $ is said to be \textit{satisfied} if $ \mu_t(i) $ is among $ i $'s most preferred objects among $ O_t $. Otherwise, $ i $ is said to be \textit{unsatisfied}.
		
		\item \textbf{Pointing}: Let each agent point to one holder of his favorite objects among the remaining ones, such that at least one cycle is generated and at least one agent in the cycle strictly prefers his pointee's object to the object he holds. These cycles are called \textit{beneficial trading cycles}.\footnote{Whenever the pointing stage is reached, a beneficial trading cycle must exist. For example, the pointing rules used by existing algorithms can generate such beneficial trading cycles.}

		\item \textbf{Trading}: Clear the beneficial trading cycles generated in the Pointing stage, so that every agent in every such cycle receives his pointee's object. Go to the next step.
	\end{itemize}

	In each step of GTTC, the pointing stage generates at least one beneficial trading cycle, so at least one agent obtains a strictly better object after trading. Since the numbers of agents and objects are finite, GTTC must stop in finitely many steps. Write the departing
	groups in order of departure as $(G_1, G_2, \ldots, G_K)$. An agent may become satisfied in some step of the algorithm but departs in a later step. Once an agent becomes satisfied, he remains satisfied in subsequent steps, until he departs with his assignment. By varying agents' pointees and the cycles generated in the pointing stage, GTTC may find different allocations. We prove that all such allocations belong to the rectified strong core.

	\begin{proposition}\label{prop:GTTC:rectifiedcore}
	Every outcome of GTTC belongs to the rectified strong core.
   \end{proposition}

	 The proof is presented in Appendix \ref{appendix:GTTC:rectifiedcore}. Below, Example \ref{example:GTTC:rectifiedstrongcore} shows that GTTC may not find all elements of the rectified strong core.

 \begin{example}[GTTC $\subsetneq$ rectified strong core]\label{example:GTTC:rectifiedstrongcore}
	Consider the following market with three agents.
	
	\begin{table}[!h]
		\centering
		\begin{subtable}[b]{.3\linewidth}
			\centering
			\begin{tabular}{cccc}
				& $ 1 $ & $ 2 $ & $ 3 $  \\ \hline
				$ \w $: & $ a $ & $ b $ & $ c $ \\ \hline
				$ \mu $: & $ a $ & $ c $ & $ b $ \\
				$ \sigma $: & $ b $ & $ a $ & $ c $\\
				$ \delta $: & $ b $ & $ c $ & $ a $ \\
				$ \eta $: & $ c $ & $ a $ & $ b $
			\end{tabular}
		\end{subtable}
		\begin{subtable}[b]{.3\linewidth}
			\centering
			\begin{tabular}{ccc}
				$ \succsim_1 $	& $ \succsim_2 $ & $ \succsim_3 $  \\ \hline
				$ b $ & $ a,c $ & $ b $ \\
				$ a,c $ & $ b $  & $ a,c $ \\
				& & \\
				& \\
				&
			\end{tabular}
		\end{subtable}
	\end{table}
	
	The rectified strong core equals $ \{\mu,\sigma,\delta,\eta\} $. However, among the four allocations, only $\mu$ and $\sigma$ are outcomes of GTTC: in step one, $ 1 $ and $ 3 $ must point to $ 2 $ and $ 2 $ must point to one of them; then, either $ \{1,2\} $ or $ \{2,3\}$ forms a cycle to exchange their endowments; after that, there is no more beneficial trading cycle.
\end{example}

In Example \ref{example:GTTC:rectifiedstrongcore}, agents share common indifferences: they all view objects $a$ and $c$ as indifferent. It belongs to a special setting we examine below in which an agent views any two objects as indifferent if and only if all agents view the two objects as indifferent. This special setting naturally arises in practice when objects have different types and each type may have multiple copies, so that agents view any two objects as indifferent if and only if they are different copies of the same type. In fact, whenever agents have common indifferences, the market can be modeled by this setting. We show that in this setting, the original TTC already produces rectified strong core allocations. This establishes a direct connection between the classical algorithm and our solution concept. In this setting, the strong core may still be empty; it is empty in Example \ref{example:GTTC:rectifiedstrongcore}.

 Formally, we consider a \textit{common indifferences} setting in which $ \mathcal{O} $ denotes a finite set of object types; for each $ x\in \mathcal{O} $, $ O_x $ denotes the set of its copies, so $ O=\cup_{x\in \mathcal{O}} O_x $; and each $ i\in I $ owns an object $ \w(i)\in O $ and has preferences $ \succsim_i $ over $ O $ such that, for every two distinct objects $ o $ and $o'$,  $ o\sim_i o' $ if and only if $ \{o,o'\}\subseteq O_x $ for some $ x\in \mathcal{O} $. In words, agents have strict preferences over object types and are indifferent among the copies of each type. Let $ I_x $ denote the set of owners of the copies of each $ x\in \mathcal{O} $.  The proof of Proposition \ref{prop:rectifiedcore:multiplecopies} is presented in Appendix \ref{appendix:exclusioncore}.

 \begin{proposition}\label{prop:rectifiedcore:multiplecopies}
 	When agents have common indifferences, all outcomes of TTC belong to the rectified strong core.
 \end{proposition}

\section{Other cooperative solutions}\label{section:othersolutions}

\subsection{Exclusion core}\label{section:exclusioncore}

By interpreting endowments as a distribution of exclusion rights, \cite{balbuzanov2019endowments} introduce the \textit{exclusion core} in a general indivisible object allocation model that subsumes the housing market model as a special case. 
In a market, a coalition directly controls its members' endowments and holds the right to evict others who occupy them. By leveraging this exclusion right, the coalition can gain indirect control of more objects along chains of occupation. The exclusion core consists of allocations in which no coalition can strictly benefit all members via evicting others from its controlled objects. Unlike the strong and weak cores, an exclusion blocking coalition is not required to allocate its own endowments. \cite{balbuzanov2019endowments} consider strict preferences and show that the exclusion core is more effective than the conventional cores in eliminating unintuitive allocations in their general model, and it coincides with the strong core in the housing market model. Given this result, it is natural to consider the exclusion core as a candidate solution for the housing market model under weak preferences.

We begin by defining the exclusion core for the housing market model. In a market $ M(\w,\succsim_I) $, given an allocation $ \mu $, every coalition $ C $ directly controls $ \w(C) $ and indirectly controls the endowments of those connected to $ \w(C) $ via chains of occupation. Formally, $C$ indirectly controls the endowment of any $i\in I\backslash C$ if there exists an agent $j\in C$ and a (possibly empty) set of agents $\{i_1,i_2,\ldots,i_K\}\subseteq I\backslash C$ such that $\mu(i)=\w(i_1)$, $\mu(i_k)=\w(i_{k+1})$ for every $k\in \{1,2,\ldots,K-1\}$, and $\mu(i_K)=\w(j)$. Let $\Omega(C|\w,\mu)$ denote the set of objects directly or indirectly controlled by $C$.\footnote{Using the formulation of \cite{balbuzanov2019endowments}, $\Omega(C|\w,\mu)=\w(\cup_{k=0}^\infty C^k)$, 
	where $ C^0=C $ and $ C^k=C^{k-1}\cup \{i\in I\backslash C^{k-1}:\mu(i)\in \w(C^{k-1})\} $ for every $ k\ge 1 $.} 

\begin{definition}[\citealp{balbuzanov2019endowments}]\label{definition:exclusion:core}
	In a market $ M(\w,\succsim_I) $, an allocation $ \mu $ is \textbf{exclusion blocked} by a coalition $ C $ via another allocation $ \sigma $ if 
	\begin{enumerate}
		\item $ \forall i\in C $, $ \sigma(i) \succ_i \mu(i) $;
		
		\item $ \forall j\in I\backslash C $, $\mu(j)\succ_j \sigma(j)\implies \mu(j)\in  \Omega(C|\w,\mu)$.
	\end{enumerate}
	The \textbf{exclusion core} consists of allocations that are not exclusion blocked.
\end{definition}

An auxiliary concept, \textit{direct exclusion core}, is defined by replacing condition (2) of Definition \ref{definition:exclusion:core} with ``$ \forall j\in I\backslash C $, $\mu(j)\succ_j \sigma(j)\implies \mu(j)\in  \w(C)$''. So, it is the set of allocations that no coalition can block by evicting others only from its directly controlled objects. 	
The direct exclusion core is a superset of the exclusion core, and both are Pareto efficient. However, the direct exclusion core may be strictly larger than the strong core in the housing market model even under strict preferences.

We find that under weak preferences, the exclusion core no longer coincides with the strong core and may also be empty. Moreover, there may exist two allocations that are indifferent for all agents, but one belongs to the exclusion core yet the other does not. Both the rectified strong core and the conventional cores are indifference-closed: two indifferent allocations either both belong to them or both do not.

\begin{customexample}{\ref{example:strongcore:empty} revisited}[$ \emptyset= $ strong core $ \subsetneq  $ exclusion core]
	Under $ \succsim'_I $, the strong core is empty, while the exclusion core equals $ \{\mu,\delta\} $. In either $ \mu $ or $ \delta $, agent $ 2 $ cannot be made strictly better off and so cannot join any exclusion blocking coalition. Then, $ 1 $ or $ 3 $ alone cannot evict the other.
\end{customexample}

\begin{example}[Exclusion core $ =\emptyset$]\label{example:exclusioncore:empty}
	Consider the following market with three agents. 
	
	\begin{table}[H]
		\centering
		\begin{subtable}[b]{.3\linewidth}
			\centering
			\begin{tabular}{cccc}
				& $ 1 $ & $ 2 $ & $ 3 $  \\ \hline
				$ \w $: & $ a $ & $ b $ & $ c $ \\ \hline
				$ \mu $: & $ b $ & $ c $ & $ a $ \\
				$ \sigma $: & $ c $ & $ a $ & $ b $
			\end{tabular}
		\end{subtable}
		\begin{subtable}[b]{.3\linewidth}
			\centering
			\begin{tabular}{ccc}
				$ \succsim_1 $	& $ \succsim_2 $ & $ \succsim_3 $  \\ \hline
				$ b $ & $ a,c $ & $ b $ \\
				$ c $ & $ b $ & $ a $ \\
				$ a $ & & $ c $\\
			\end{tabular}
		\end{subtable}
	\end{table}
	
	In any individually rational and Pareto efficient allocation, all three agents must trade in a cycle, yielding either $ \mu $ or $ \sigma $. In either allocation, every agent controls all objects. So, $ 3 $ can exclusion block $ \mu $ via $ \sigma $, and $ 1 $ can exclusion block $ \sigma $ via $ \mu $. Hence the exclusion core is empty.
\end{example}

\begin{customexample}{\ref{example:GTTC:rectifiedstrongcore} revisited}[Exclusion core is not indifference-closed]
	The exclusion core equals $ \{\mu,\sigma\} $. Although $ \sigma $ and $ \delta $ are indifferent, $ \delta $ is exclusion blocked by $ 3 $ via $ \mu $. Similarly, although $ \mu $ and $ \eta $ are indifferent, $ \eta $ is exclusion blocked by $ 1 $ via $ \sigma $.
\end{customexample}

We identify a necessary condition and a separate sufficient condition for an allocation to belong to the exclusion core (Lemma \ref{lemma:exclusion:core:nonempty} in Appendix \ref{appendix:exclusioncore}). The necessary condition implies that if the exclusion core is nonempty, every allocation within it must be an outcome of TTC. The sufficient condition implies that once the strong core is nonempty, every strong core allocation belongs to the exclusion core. Together with Example \ref{example:strongcore:empty}, in which the exclusion core is nonempty while the strong core is empty, it means that the exclusion core is more often nonempty than the strong core. Proposition \ref{prop:exclusioncore} further shows that the exclusion core is always a subset of the rectified strong core. Since the rectified strong core coincides with the strong core whenever the latter is nonempty, it implies that the exclusion core also coincides with the strong core whenever the latter is nonempty.

\begin{proposition}\label{prop:exclusioncore}
	The exclusion core is a (possibly empty) subset of the rectified strong core and a superset of the strong core. Whenever the strong core is nonempty, the three cores coincide.
\end{proposition}

 Proposition \ref{prop:exclusioncore:multiplecopies}  characterizes the exclusion core in the common indifferences setting discussed in Section \ref{section:GTTC}: the exclusion core is nonempty and coincides with the set of TTC outcomes. Together with Proposition \ref{prop:exclusioncore}, it implies Proposition \ref{prop:rectifiedcore:multiplecopies}. The proofs are presented in Appendix \ref{appendix:exclusioncore}.

\begin{proposition}\label{prop:exclusioncore:multiplecopies}
	When agents have common indifferences, the exclusion core coincides with the set of TTC outcomes.
\end{proposition}

\subsection{The von Neumann-Morgenstern stable set}

An allocation $\sigma$ is said to \textit{weakly (strongly) dominate} another allocation $\mu$ if $\mu$ is weakly (strongly) blocked by some coalition via $\sigma$. A vNM stable set is then defined as follows.

\begin{definition}[\citealp{vonNeumann1944}]\label{definition:vNM:stable}
		In a market $ M(\w,\succsim_I) $, a subset of allocations $ A\subseteq \a(M) $ is a \textbf{vNM stable set based on weak (strong) domination}  if it satisfies the following two conditions:
		\begin{enumerate}
			\item Internal stability: Every $ \mu\in A $ is not weakly (strongly) dominated by any other $ \sigma\in A$.
			
			\item External stability: Every $ \sigma\notin A$ is weakly (strongly) dominated by some $\mu\in A$.
		\end{enumerate}
	\end{definition}
A vNM stable set based on weak domination is Pareto efficient if it exists, yet a vNM stable set based on strong domination may contain Pareto inefficient allocations. \cite{wako2007nonexistence} shows that a vNM stable set based on either strong or weak domination generally does not exist; examples are available in his paper. 

A vNM stable set may not be unique when it exists. In Example \ref{example:strongcore:empty}, under $\succsim'_I$, both $\{\mu\}$ and $\{\delta\}$ are weak domination vNM stable sets.  \cite{wako1991some} shows that when the strong core is nonempty, the strong core is the unique vNM stable set based on weak domination. Several papers modify the definition of weak domination and prove that the corresponding vNM stable set exists and coincides with the set of competitive allocations, which may contain Pareto inefficient allocations and may be strictly larger than a nonempty strong core; see the discussions in the related literature part of the introduction.

\subsection{Myopic stable set} \cite{demuynck2019myopic} introduce the \textit{myopic stable set} (MSS) for coalition formation in a general class of social environments. Compared to the vNM stable set, the main advantages of MSS are its existence in any social environment and its uniqueness in social environments with finite states.  
The housing market model is an environment with finite states, because there are only finitely many possible allocations.

\begin{definition}[\citealp{demuynck2019myopic}]\label{definition:myopic:stable}
		In a market $ M(\w,\succsim_I) $, a subset of allocations $ A\subseteq \a(M) $ is a \textbf{myopic stable set based on weak domination}  if it satisfies three conditions:
		\begin{enumerate}
			\item Deterrence of external deviations: Every $ \mu\in A $ is not weakly dominated by any $ \sigma\notin A$.
			
			\item Iterated external stability: For every $\sigma\notin A$, there exist an integer $K\ge 1$ and distinct allocations $\mu_0,\mu_1,\ldots,\mu_K$ such that $\mu_0=\sigma$, $\mu_K\in A$, and $\mu_{k-1}$ is weakly dominated by $\mu_k$ for every $k=1,\ldots,K$.

			\item Minimality: No strict subset $A'\subsetneq A$ satisfies the conditions (1) and (2).
		\end{enumerate}
\end{definition}

In their Online Appendix A.5, \cite{demuynck2019myopic} prove that in the housing market model under strict preferences, MSS based on weak domination coincides with the strong core. We show that under weak preferences, however, the weak domination MSS may contain Pareto inefficient allocations. See the following example.

\begin{customexample}{\ref{example:exclusioncore:empty} revisited}[MSS contains Pareto inefficient allocations]
	In Example \ref{example:exclusioncore:empty}, $ \mu $ and $ \sigma $ are Pareto efficient allocations. Besides $ \w $, there are three other possible allocations: $ \mu'(1,2,3)=(a,c,b) $,  $ \sigma'(1,2,3)=(b,a,c) $, and $ \delta(1,2,3)=(c,b,a) $. The weak domination MSS equals $ \{\mu,\mu',\sigma,\sigma'\} $, which contains the Pareto inefficient allocations $ \mu' $ and $ \sigma' $.
	
	To verify that this set is an MSS, we draw the following graph to represent the domination relationships between allocations. In the graph, an allocation points to another one if it is weakly dominated by the latter. The four allocations within the dotted circle in the graph form a cycle in domination relationships, and they do not point to any allocation outside the circle. Every allocation outside the circle points to at least one allocation within the cycle. Thus, the four allocations within the circle constitute the unique MSS.

	\begin{center}
		\begin{tikzpicture}[
			xscale=0.5, yscale=0.5,
			->, >=stealth, thick,
			inner/.style={circle, draw=red, text=red, minimum size=0.6cm, inner sep=1pt},
			outer/.style={circle, draw,           minimum size=0.6cm, inner sep=1pt}
		]
			\node[inner] (mu)   at (  0,   2.5) {$\mu$};
			\node[inner] (mup)  at (  2.5,  0 ) {$\mu'$};
			\node[inner] (sig)  at (  0,  -2.5) {$\sigma$};
			\node[inner] (sigp) at ( -2.5,  0 ) {$\sigma'$};
			\draw[dashed, red] (0,0) circle (3.3cm);
			\node[outer] (delta) at ( 5,    0  ) {$\delta$};
			\node[outer] (omega) at (-5,    0  ) {$\omega$};

			\draw[red] (mu)   -- (mup);
			\draw[red] (mup)  -- (sig);
			\draw[red] (sig)  -- (sigp);
			\draw[red] (sigp) -- (mu);
			\draw[red] (mup)  to[bend left=25] (sigp);   
			\draw[red] (sigp) to[bend left=25] (mup);    

			\draw (delta) -- (mu);
			\draw (delta) -- (sig);

			\draw (omega) to[out=-20, in=-160] (delta);
			\draw (omega) to[out=40,  in=200] (mu);
			\draw (omega) to[out=35,  in=145] (mup);
			\draw (omega) to[out=-40, in=160] (sig);
			\draw (omega) -- (sigp);
		\end{tikzpicture}
	\end{center}

\end{customexample}

Both the vNM stable set and the MSS are setwise solutions; whether an allocation belongs to the set depends on which other allocations are in the set. In contrast, the core concepts we discuss are pointwise solutions; whether an allocation belongs to a core is a property of the allocation, irrespective of other allocations in the core.

\subsection{Bargaining set} 

\cite{yilmaz2022stability} apply the concept of \textit{bargaining set} originally developed by \cite{aumann1964bargaining} to the housing market model. Using a farsightedness argument,  they treat a blocking as legitimate only if there does not exist a counter-blocking. An allocation belongs to the bargaining set if it belongs to the weak core and whenever it is weakly blocked by a coalition via a new allocation, there exists a counter-blocking coalition that overlaps with the original coalition and that blocks the new allocation by claiming their welfare in the original allocation.

\begin{definition}[\citealp{yilmaz2022stability}]\label{definition:bargainingset}
		In a market $ M(\w,\succsim_I) $, an allocation $ \mu$ belongs to the \textbf{bargaining set}  if it satisfies the following two conditions:
		\begin{enumerate}
			\item It is not strongly blocked by any coalition.
			
			\item If it is weakly blocked by a coalition $C$ via another $\sigma$, then there exists a coalition $ C' $ that weakly blocks $ \sigma $ via some $ \mu' $ such that $ C'\cap C\neq \emptyset $ and for all $ i\in C' $, $ \mu'(i)\sim_i \mu(i) $.
		\end{enumerate}
	\end{definition}

	In the above definition, when a coalition $C$ is considered to weakly block $\mu$ via another $\sigma$, since the usual definition of weak blocking does not restrict the assignments for the agents outside $C$ under $\sigma$, \citeauthor{yilmaz2022stability} select $\sigma$ in which  all agents outside $C$ who are involved in cycles along with members of $ C $ under $ \mu $ receive their own endowments under $ \sigma $, while those who are unaffected retain their assignments under $\mu$.

The bargaining set is nonempty, Pareto efficient, and lies between the strong core and the weak core. However, we note that the bargaining set may be strictly larger than the strong core when the latter is nonempty; see Example \ref{example:bargainingset} in Appendix \ref{appendix:examples}. Since the rectified strong core coincides with a nonempty strong core, it means that the bargaining set is not a subset of the rectified strong core. Nor is the rectified strong core a subset of the bargaining set; see Example \ref{example:bargainingset:2} in Appendix \ref{appendix:examples}. 

Cores are myopic solutions, whereas the bargaining set is a farsighted solution. Consequently, verifying the bargaining set requires a two-step process: for every potential coalition blocking, one must verify the existence of a potential counter-blocking coalition, which could be computationally more challenging than verifying a core concept.

\section{Conclusion}\label{section:conclusion}

This paper studies core concepts in the Shapley-Scarf housing market model under unrestricted weak preferences. The two standard core concepts exhibit well-known deficiencies: the strong core may be empty, and the nonempty weak core may contain Pareto inefficient allocations. We introduce the \textit{rectified strong core} by adding a single \textit{no outside option} condition to the weak blocking requirement: an unaffected agent may join a blocking coalition only if every object he views as indifferent to his current assignment is owned by the coalition. Under strict preferences this condition holds automatically. The rectified strong core can be viewed as the natural extension of the strong core to weak preferences. It is always nonempty and Pareto efficient, lies between the weak and strong cores, and coincides with the strong core whenever the latter is nonempty. 

We justify the no outside option condition from two angles. The behavioral argument shows that the condition unifies two interpretations for unaffected agents' participation: a \textit{compellability} interpretation, under which an unaffected agent who relies on strictly-better-off members' endowments to maintain his welfare can be viewed as compelled to participate, and a \textit{neutrality} interpretation, under which a group of self-sufficient unaffected agents would displace exactly the same agents outside the coalition if they reclaim their own endowments, so their participation leaves the external impact unchanged. The tightness argument shows that any relaxation of the condition can render the corresponding core empty. To find rectified strong core allocations, we introduce the generalized top trading cycles (GTTC) algorithm and prove that all of its outcomes belong to the rectified strong core; in housing markets with common indifferences, the original TTC already suffices. Among alternative solutions, the exclusion core of \cite{balbuzanov2019endowments} is always a subset of the rectified strong core but may be empty; the von Neumann-Morgenstern stable set may not exist; the myopic stable set of \cite{demuynck2019myopic} may contain Pareto inefficient allocations; and the bargaining set of \cite{yilmaz2022stability} is nonempty and Pareto efficient but may be strictly larger than a nonempty strong core.

Several directions remain open. A characterization of the rectified strong core, whether axiomatic, structural, or algorithmic, is a natural next step; for instance, understanding which allocations belong to the rectified strong core and how its structure depends on the preference profile may yield further insight into the solution concept. More broadly, the no outside option condition may be of independent interest beyond the housing market model: extending the analysis to richer environments such as two-sided matching markets, general exchange economies, or models with priorities may shed further light on the role of unaffected agents in cooperative solution concepts under weak preferences.

	\newpage

	\appendix
	
	\section*{Appendix}

\section{Proof of Theorem \ref{theorem:rectifiedcore}}\label{appendix:Theorem1}

	(\textbf{Nonemptiness}) Nonemptiness follows from Proposition~\ref{prop:GTTC:rectifiedcore}, proved independently in Appendix~\ref{appendix:GTTC:rectifiedcore}. 
	
	(\textbf{Pareto efficiency}) Every Pareto inefficient allocation is rectification blocked by the grand coalition $ I $ via a Pareto improvement.

	(\textbf{Relationships with strong core and weak core}) The rectified strong core is a subset of the weak core, because any strong blocking is a rectification blocking. 	
	It remains to prove that the rectified strong core coincides with the strong core whenever the strong core is nonempty. The proof utilizes the condition provided by \cite{quint2004houseswapping} that characterizes when the strong core is nonempty. Below, we first introduce their condition and then prove our result.

	Given any nonempty $ I'\subseteq I $,  a nonempty $ J\subseteq I' $ is called a \textbf{minimal self-mapped set in $ I' $} if $ \cup_{i\in J}B_i(\w(I'))= \w(J) $ and no strict nonempty subset $ J' \subsetneq J $ satisfies $ \cup_{i\in J'}B_i(\w(I'))= \w(J') $. In words, for every $ i\in J $, his most preferred objects among $ \w(I') $ are owned by $ J $ and his own endowment is most preferred by some agent in $ J $ among $ \w(I') $, but no strict subset of $ J $ satisfies this condition.
  Let $ T^*=(T_1,T_2,\ldots,T_{t^*}) $ denote a partition of $ I $; that is, every $ T_k\in T^* $ is a nonempty subset of $ I $, any two distinct $ T_k,T_{k'}\in T^* $ are disjoint, and  $ \bigcup_{k=1}^{t^*}T_k=I $. $T^*$ is a \textbf{top trading segmentation} (TTS) if, for every $ T_k\in T^* $, $ T_k $ is a  minimal self-mapped set in $ \bigcup_{\ell=k}^{t^*}T_\ell $, and there exists a one-to-one mapping $ \mu_k $ from $ T_k $ to $ \w(T_k) $ such that, for every $ i\in T_k $, $ \mu_k(i)\in B_i\big(O\backslash \bigcup_{\ell=1}^{k-1}\w(T_\ell) \big) $. In words,  each $ T_k $ can distribute its endowments among its members so that each member obtains a most preferred object among all objects owned by $ \bigcup_{\ell=k}^{t^*}T_\ell $.
	
	The existence of a TTS is necessary and sufficient for the strong core to be nonempty.

	\begin{lemma}[\citealp{quint2004houseswapping}]\label{lemma:strong:core:nonempty}
		In any market $ M(\w,\succsim_I) $, 
		the strong core is nonempty if and only if there exists a TTS $ T^*=(T_1,T_2,\ldots,T_{t^*}) $. When the strong core is nonempty, an allocation $ \mu $ belongs to the strong core if and only if, for every $ T_k\in T^* $ and every $ i\in T_k$, $ \mu(T_k)=\w(T_k) $ and $ \mu(i)\in B_i(O\backslash \bigcup_{\ell=1}^{k-1}\w(T_\ell)) $.
	\end{lemma}

	When the strong core is nonempty, there exists a TTS $ T^*=(T_1,T_2,\ldots,T_{t^*}) $. 	
	Let $ \mu $ be an allocation in the rectified strong core. Below, we prove that for every $ T_k\in T^* $ and every $ i\in T_k $, $ \mu(i)\in B_i\big(O\backslash \bigcup_{\ell=1}^{k-1}\w(T_\ell) \big) $. Then, by Lemma \ref{lemma:strong:core:nonempty}, $ \mu $ belongs to the strong core. 
	
	We first prove that, for every $ i\in T_1 $, $ \mu(i)\in B_i(O) $. Suppose it is not true. Let $ \mu_1 $ be a one-to-one mapping  from $ T_1 $ to $ \w(T_1) $ such that, for every $ i\in T_1 $, $ \mu_1(i)\in B_i(O) $. Then, $ T_1 $ can rectification block $ \mu $ via an allocation $ \mu' $ in which, for every $ i\in T_1 $, $ \mu'(i)=\mu_1(i) $. Condition (1) and (2) of Definition \ref{definition:rectified:core} are obviously satisfied. The key is to verify condition (3). It is satisfied because all of the most preferred objects for each agent in $ T_1 $ are owned by $ T_1 $. 
	
	Similarly, if it is not true that, for every $ i\in T_2 $, $ \mu(i)\in B_i(O\backslash \w(T_1)) $, we then let $ \mu_2 $ be a one-to-one mapping from $ T_2 $ to $ \w(T_2) $ such that, for every $ i\in T_2 $, $ \mu_2(i)\in B_i(O\backslash \w(T_1)) $. Then, $ T_1\cup T_2 $ can rectification block $ \mu $ via an allocation $ \mu' $ in which, for every $ i\in T_1 $, $ \mu'(i)=\mu(i) $, and for every $ i\in T_2 $, $ \mu'(i)=\mu_2(i) $. The key is to verify condition (3) of Definition \ref{definition:rectified:core}. The condition is satisfied because all of the most preferred objects for each agent in $ T_1 $ are owned by $ T_1 $, and all of the most preferred objects for each agent in $ T_2 $ among $ O\backslash \w(T_1) $ are owned by $ T_2 $.

	The above argument inductively holds for every remaining $ T_k \in T^*$; that is, we can prove that, for every $ i\in T_k $,  $ \mu(i)\in B_i\big(O\backslash \bigcup_{\ell=1}^{k-1}\w(T_\ell) \big) $. So, $ \mu $ belongs to the strong core.

	Since the rectified strong core is always a superset of the strong core, it means that whenever the strong core is nonempty, it coincides with the strong core.

	\section{Proof of Proposition \ref{prop:participation:tightness}}\label{appendix:proof:tightness}	

  In any housing market, let $I=\{1,2,\ldots,n\}$ and $O=\{o_1,o_2,\ldots,o_n\}$, with agent $i$ owning $o_i$. Suppose that there exist $X\subseteq O$ and $Z\subsetneq X$ with $ p(X,Z)=1 $. We then construct a preference profile and show that the $p$-core is empty. So, the $p$-core is always nonempty only if $p$ is precisely the participation function defining the rectified strong core. Since the rectified strong core is always nonempty, we obtain the ``if and only if'' result.

	Let $|X|=k $ and $|Z|=k'$. So, $0<k'<k\le n$. We construct a preference profile $\succsim_I$ below, analogous to $ \succsim'_I $ in Example \ref{example:strongcore:empty}.  We show that the $p$-core is empty under $\succsim_I$.

	\begin{table}[H]
		\centering
		\begin{tabular}{cccc}
			$ \succsim_1 $	& $ \succsim_2 $ & $ \succsim_3 $ & $ \succsim_{i: i\ge 4} $ \\ \hline
			$ o_2 $ & $ o_1,o_2,\ldots,o_k $ & $ o_2 $ & $o_i$ \\
			$ o_1 $ & $\vdots$ & $ o_3 $ & $\vdots$ \\
			$ \vdots $ & & $ \vdots $\\
			& \\
		\end{tabular}
	\end{table}
	
	Since every agent $i\ge 4$ most prefers his own endowment, any allocation in the $p$-core must assign $o_i$ to $i$ for all $i\ge 4$. Then,  since 1 and 3 most prefer 2's endowment and then their own, while 2 is indifferent among the endowments of the first $k$ agents, there are only three candidates for the $p$-core: $\mu$, in which agents $1$ and $2$ exchange endowments and the others receive their own endowments;  $\sigma$,  in which $2$ and $3$ exchange endowments and the others receive their own endowments; and the endowment allocation $\w$. 
	
	However, $\mu$ is $p$-blocked by $C=\{2,3,4,\ldots, k'+1\}$ via $\sigma$. Here $C_{\sigma\sim\mu}=\{2,4,\ldots,k'+1\}$. For each $i\in C_{\sigma\sim\mu}$ with $i\ge 4$, full-coverage sufficiency guarantees their admissibility to the coalition. For agent $2$, we have $|\mathcal{I}_2(\mu(2))\cap \w(C)|=k'$. So, anonymity guarantees his admissibility to the coalition. 
	By symmetric arguments,  $\sigma$ is $p$-blocked by $C'=\{1,2,4,\ldots, k'+1\}$ via $\mu$. Since $\w$ is not Pareto efficient, it is $p$-blocked by $C$ via $\sigma$ or by $C'$ via $\mu$. Therefore, the $p$-core is empty.

\section{Proof of Proposition \ref{prop:GTTC:rectifiedcore}}\label{appendix:GTTC:rectifiedcore}

Let $ \mu $ be any outcome of GTTC. We prove that $ \mu $ belongs to the rectified strong core. 

We first prove that $ \mu $ is Pareto efficient. Let $ (G_1,G_2,\ldots,G_K) $ be the order of the departing groups in the procedure of GTTC that generates $ \mu $. Every member of $ G_1 $ obtains one of his most preferred objects among $ O $. So, they cannot be made strictly better off. All of their most preferred objects are also held by $ G_1 $. After $ G_1 $ departs with their assignments, every member of $ G_2 $ obtains one of his most preferred objects among $ O\backslash \mu(G_1) $. Thus, they cannot be made strictly better off without making any member of $ G_1 $ worse off. Applying this argument inductively to the remaining groups, we conclude that $ \mu $ is Pareto efficient.

We then prove that $\mu$ is unblocked. Since $ \mu $ is Pareto  efficient, it cannot be rectification blocked by $ I $. Below, we prove that $ \mu $ cannot be rectification blocked by any $ C\subsetneq I $.

Suppose that $ \mu $ is rectification blocked by a coalition $ C $ via another $ \sigma $. Without loss of generality, we choose $\sigma$ such that, for every $i\in I\backslash C$, $\mu(i)\neq \sigma(i)$ if and only if $\mu(i)\in \w(C)$. Among the agents in $ C_{\sigma>\mu} $, let $ j^* $ be an agent who first becomes satisfied in the procedure of GTTC. If there are multiple agents who become satisfied simultaneously, let $ j^* $ be one of them. Since $ \sigma(j^*)\succ_{j^*} \mu(j^*) $, $ \sigma(j^*) $ must be removed in the algorithm before $ j^* $ becomes satisfied. Let $ (G_1,G_2,\ldots,G_L) $ be the order of groups that depart before $j^*$ becomes satisfied. Then, we have $ \sigma(j^*)\in \mu(G_1\cup G_2\cup \cdots \cup G_L) $.

We first prove that there exists $i^*\in G_1\cup G_2\cup \cdots \cup G_L$ such that $ \mu(i^*)\succ_{i^*} \sigma(i^*) $. Since $ \sigma(j^*)\in \mu(G_1\cup G_2\cup \cdots \cup G_L) $ and $ j^*\notin G_1\cup G_2\cup \cdots \cup G_L $, there must exist $i^*\in G_1\cup G_2\cup \cdots \cup G_L$ such that $ \sigma(i^*)\notin \mu(G_1\cup G_2\cup \cdots \cup G_L) $. However, since for every $ i\in G_1\cup G_2\cup \cdots \cup G_L $, $ \{o\in O:o\succsim_i \mu(i)\}\subseteq \mu(G_1\cup G_2\cup \cdots \cup G_L)  $, it must be that $ \mu(i^*)\succ_{i^*} \sigma(i^*) $. Therefore, we have $i^*\in I\backslash C$ and, by our selection of $\sigma$,  $\mu(i^*)\in \w(C)$. 

Below we prove a claim. Applying the claim, we obtain a contradiction that $i^*\in C$.

\begin{claim}\label{claim:3}
	In GTTC, for any beneficial trading cycle that is generated before $j^*$ becomes satisfied,  if an agent in the cycle belongs to $ C $, then all agents in the cycle belong to $ C $. 
\end{claim}

\begin{proof}[Proof of Claim \ref{claim:3}]
	
	In GTTC, let $ (Y_1,Y_2,\ldots,Y_K) $ denote the order of cycles that are generated before $j^*$ becomes satisfied and that involve an agent from $ C $. Here $ Y_k $ denotes the set of agents involved in the corresponding cycle.  If several cycles are generated in the same step, their relative ranking is arbitrary in the order.
	
	Without loss of generality, we represent the first cycle $ Y_1 $ by
	\[
	i_1 \rightarrow  i_2 \rightarrow i_3 \rightarrow \cdots \rightarrow i_k  \rightarrow i_1,
	\]
	where $ i_\ell\rightarrow i_{\ell+1} $ means that after clearing the cycle, $ i_\ell $ obtains the object held by $ i_{\ell+1} $. Suppose that $ i_1\in  C$. Let $ o_2 $ be the object held by $ i_2 $. After clearing the cycle, $ i_1 $ obtains $o_2$. So, $ o_2\sim_{i_1} \mu(i_1) $. 	
	Because $ Y_1 $ is generated before $j^*$ becomes satisfied and $j^*$ is the first agent among $C_{\sigma>\mu}$ who becomes satisfied, we have $ i_1\in C_{\sigma\sim \mu}  $. Therefore, $\mathcal{I}_{i_1} (\mu(i_1))\subseteq \w(C) $. This means that the owner of $o_2$ must belong to $ C $. If $ i_2 $ is not the owner of $o_2$, then the owner of $ o_2 $ must be involved in a cycle before $ Y_1 $ is generated. However, this contradicts the definition of $ Y_1 $. So, $ i_2 $ must be the owner of $o_2$. Then, similar to $ i_1 $, it must be that $ i_2\in C_{\sigma\sim \mu} $. By the same argument, we can show that $i_3\in C_{\sigma\sim \mu}$. Applying the argument inductively to the remaining agents in $Y_1$, we conclude that they all belong to $ C_{\sigma\sim \mu} $.
	
	We then consider the second cycle $Y_2$. For convenience, we still represent it by
	\[
	i_1 \rightarrow  i_2 \rightarrow i_3 \rightarrow \cdots \rightarrow i_k  \rightarrow i_1,
	\]
	and assume that $ i_1\in  C$. Let $ o_2 $ be the object held by $ i_2 $.  After clearing the cycle, $ i_1 $ obtains $o_2$.  Because $ Y_2$ is generated before $j^*$ becomes satisfied and $j^*$ is the first agent among $C_{\sigma>\mu}$ who becomes satisfied, it must be that $ i_1\in C_{\sigma\sim \mu}  $. Therefore, $\mathcal{I}_{i_1} (\mu(i_1))\subseteq \w(C) $. So, the owner of $o_2$ must belong to $ C$. Then there are two cases. 
	
	If $ i_2 $ is not the owner of $o_2$, then the owner of $o_2$ and $ i_2 $ must be involved in cycles that are generated earlier than $ Y_2 $, and after clearing these cycles, $i_2$ obtains $o_2$. Since $ Y_1 $ is the only cycle before $Y_2$ that involves an agent from $C$, the owner of $o_2$ and $ i_2 $ must be involved in $ Y_1 $. Then, by the above argument for $Y_1$, $ i_{2}\in C_{\sigma\sim \mu} $. 
	
	If $ i_2 $ is the owner of $o_2$, then $i_2\in C$. Similar to $ i_1 $, we have $i_2\in C_{\sigma\sim \mu} $. 
	
	So, we always have $i_2\in C_{\sigma\sim \mu} $. Applying the argument inductively to the remaining agents in $Y_2$, we conclude that all agents in $Y_2$ belong to $ C_{\sigma\sim \mu} $.
	
	The above argument  inductively applies to all cycles.
\end{proof}

Let $i^\circ$ be the owner of $\mu(i^*)$; that is, $ \w(i^\circ)= \mu(i^*)$.  Since $\mu(i^*)\in \w(C)$, $i^\circ\in C$. Now,  since $i^*$ departs with $\mu(i^*)$  at some point before $j^*$ becomes satisfied, $ i^* $ and $ i^\circ $ must be respectively involved in a sequence of cycles such that an agent $ i_1 $ first obtains $ \w(i^\circ) $ from $ i^\circ $ in a cycle $Z_1$, then an agent $ i_2 $ obtains $ \w(i^\circ) $ from $ i_1 $ in a cycle $Z_2$, and so on, until $ i^* $ obtains $ \w(i^\circ) $ from an agent $ i_k $ in a cycle $Z_{k+1}$. By Claim \ref{claim:3}, since $ i^\circ\in C $, all agents in $ Z_1 $ belong to $ C $. So, $ i_1\in C $. Again, by Claim \ref{claim:3}, all agents in $ Z_2 $ belong to $ C $. So, $ i_2\in C $. By applying Claim \ref{claim:3} inductively to all these cycles, we conclude that all agents in these cycles belong to $ C $. So, $ i^*\in C $. However, this contradicts $i^*\in I\backslash C$. 

\section{Proofs of Proposition \ref{prop:rectifiedcore:multiplecopies}, \ref{prop:exclusioncore}, and \ref{prop:exclusioncore:multiplecopies}}\label{appendix:exclusioncore}

We first introduce Lemma \ref{lemma:exclusion:core:nonempty}, which provides a necessary condition and a separate sufficient condition for an allocation to belong to the exclusion core. Lemma \ref{lemma:exclusion:core:nonempty} is used in the proofs of Proposition \ref{prop:exclusioncore} and  Proposition \ref{prop:exclusioncore:multiplecopies}, which together imply Proposition \ref{prop:rectifiedcore:multiplecopies}.
	
	\begin{lemma}\label{lemma:exclusion:core:nonempty}
	In any market $ M(\w,\succsim_I) $:
	\begin{enumerate}
		\item An allocation $ \mu $ belongs to the exclusion core \textbf{only if} $ \mu $ is Pareto efficient and there exists a partition of agents $ T=(T_1,T_2,\ldots,T_t) $ such that, for every $ T_k\in T $ and every $ i\in T_k $, $\mu(T_k)=\w(T_k) $ and $ \mu(i)\in B_i\big(O\backslash \bigcup_{\ell=1}^{k-1}\w(T_\ell) \big) $.
		
		\item An allocation $ \mu $ belongs to the exclusion core \textbf{if} $ \mu $ is Pareto efficient and there exists a partition of agents $ T=(T_1,T_2,\ldots,T_t) $ such that, for every $ T_k\in T $ and every $ i\in T_k $, $\mu(T_k)=\w(T_k) $ and $ \mu(i)\in B_i\big(O\backslash \bigcup_{\ell=1}^{k-1}\w(T_\ell) \big) $, and, additionally, for any allocation $ \sigma $ with $ I_{\sigma>\mu}\neq \emptyset $, there exists $ T_k\in T $ such that $ T_k \cap I_{\mu>\sigma}\neq \emptyset$ and $ k<k' $ for every $ T_{k'}\in T $ with $ T_{k'} \cap I_{\sigma>\mu}\neq \emptyset$.
 	\end{enumerate} 
\end{lemma}

The partition $ T=(T_1,T_2,\ldots,T_t) $ in Lemma \ref{lemma:exclusion:core:nonempty} may not be a TTS, the existence of which is necessary and sufficient for the strong core to be nonempty (Lemma \ref{lemma:strong:core:nonempty} in Appendix \ref{appendix:Theorem1}). Specifically, each $ i\in T_k $ can receive a most preferred object in $ O\backslash \bigcup_{\ell=1}^{k-1}\w(T_\ell) $ from $ \w(T_k) $, but $ i $'s most preferred objects  need not be owned exclusively by $ T_k $.  

The ``only if'' result implies that every element of a nonempty exclusion core must be an outcome of TTC.\footnote{The condition $\mu(T_k)=\w(T_k) $ means that the agents in each $ T_k $ exchange their endowments. Hence $ \mu $ can be produced by TTC via generating cycles in an order consistent with $(T_1,T_2,\ldots,T_t) $.} In fact, the proof below directly shows this.

	\begin{proof}[Proof of Lemma \ref{lemma:exclusion:core:nonempty}]
		(\textbf{Only if})  Suppose that the exclusion core is nonempty and $ \mu $ is its element. So, $ \mu $ is Pareto efficient. In $ \mu $, agents can be partitioned into disjoint groups such that the agents in each group trade their endowments in a cycle. Without loss of generality, represent the cycle by 
		\[
		i_1 \rightarrow  i_2 \rightarrow i_3 \rightarrow \cdots \rightarrow i_k  \rightarrow i_1,
		\] 
		where $ i_\ell\rightarrow i_{\ell+1} $ means that $ \mu(i_\ell)=\w(i_{\ell+1}) $.\footnote{If an agent obtains his own endowment, he forms a cycle with himself.} Denote the set of agents in a typical group by $ T_k $. Therefore, for every $ T_k $, $\mu(T_k)=\w(T_k) $, and no strict subset of $ T_k $ satisfies this condition. We prove that these groups can be arranged into an order $ T=(T_1,T_2,\ldots,T_t) $ such that, for every $ T_k\in T $ and every $i\in T_k$, $ \mu(i)\in B_i\big(O\backslash \bigcup_{\ell=1}^{k-1}\w(T_\ell) \big) $.

		First, for every $ T_k\in T $, we prove that every $ i\in T_k $ most prefers his assignment $ \mu(i) $ among all objects in $ \w(T_k) $. Because the agents in $ T_k $ trade their endowments in a cycle represented above, every agent in $ T_k $ controls all objects in $ \w(T_k) $. Then, if any $ i\in T_k $ strictly prefers some object in $ \w(T_k) $ over $ \mu(i) $, then $ i$ would be able to exclusion block $ \mu $ by assigning himself the preferred object and assigning his assignment $ \mu(i) $ to the agent who receives his preferred object under $ \mu $. This is a contradiction. So, every $ i\in T_k$ must most prefer $ \mu(i) $ among all objects in $ \w(T_k) $.
		
		Second, we prove that there must exist a group in which all agents most prefer their assignments among all objects. We label this group $ T_1 $. To find $ T_1 $, start with any group $ T_a $. If some $ i_a\in T_a $ strictly prefers an object owned by another group $ T_b$ over his assignment $ \mu(i_a) $, we then examine $ T_b $. If some $ i_b\in T_b $ strictly prefers an object owned by another group $ T_c $ over his assignment $ \mu(i_b) $, we then examine $ T_c $. Continuing this search process, because there are finite groups, we must either find a group in which all agents most prefer their assignments among all objects, or find a group in which some $ i $ strictly prefers an object owned by a group we have examined over his assignment $ \mu(i) $. In the former case, we label the group $ T_1 $. We prove that the latter case is impossible. In the latter case, there exists a sequence of groups $ T'=(T'_1,T'_2,\ldots,T'_m) $ such that, in every $ T'_k\in T'\backslash T_m $, some $ i_k$ strictly prefers the object owned by some agent in $ T'_{k+1} $ over his assignment $ \mu(i_k) $, while some $ i_m\in T'_m $ strictly prefers the object owned by some agent in $ T'_1 $ over his assignment $ \mu(i_m) $. Then, $ \{i_1,i_2,\ldots,i_m\} $ can form a coalition to exclusion block $ \mu $, because these agents control all objects in $ \cup_{T_k\in T'} \w(T_k) $. This is a contradiction. See the illustration in Figure \ref{figure:proofillustration}.
		
		\begin{figure}
			\centering
			\begin{tikzpicture}[semithick,bend angle=20,xscale=1,yscale=1]
				\node (i1) at (1,0) [agent] {$1$};
				\node (i2) at (3,0) [agent] {$2$};
				\node (i3) at (2,-1.2) [agent] {$3$};

				\draw[-latex] (i1)  to (i2);
				\draw[-latex] (i2)  to (i3);
				\draw[-latex] (i3)  to (i1);

				\node (i4) at (4,2) [agent] {$4$};
				\node (i5) at (6,2) [agent] {$5$};
				\node (i6) at (5,0.8) [agent] {$6$};

				\draw[-latex] (i4)  to (i5);
				\draw[-latex] (i5)  to (i6);
				\draw[-latex] (i6)  to (i4);

				\node (i7) at (7,0) [agent] {$7$};
				\node (i8) at (9,0) [agent] {$8$};
				\node (i9) at (8,-1.2) [agent] {$9$};

				\draw[-latex] (i7)  to (i8);
				\draw[-latex] (i8)  to (i9);
				\draw[-latex] (i9)  to (i7);
				
				\draw[->,dashed] (i2) to (i4);
				\draw[->,dashed] (i5) to (i7);
				\draw[->,dashed] (i9) to (i3);
			\end{tikzpicture}
			\caption{\footnotesize Three groups $\{1,2,3\}$, $\{4,5,6\}$, and $\{7,8,9\}$ each trade endowments in a cycle (solid arrows), with every member most preferring his assignment among his group's endowments. However, agent $2$ strictly prefers $4$'s endowment, $5$ strictly prefers $7$'s endowment, and $9$ strictly prefers $3$'s endowment (dashed arrows). Since $\{2,5,9\}$ indirectly controls all objects across the three cycles, it can exclusion block the allocation by assigning each member his preferred object.}\label{figure:proofillustration}
		\end{figure}	 
		
		After finding $ T_1 $, we can repeat the above argument to find the group $ T_2 $ in which all agents most prefer their assignments among  $ O\backslash \w(T_1) $. Inductively applying the above argument, we can find the desired order of groups, $ T=(T_1,T_2,\ldots,T_t) $.

		(\textbf{If}) Suppose that $ \mu $ is a Pareto efficient allocation and there exists a partition of agents $ T=(T_1,T_2,\ldots,T_t) $ such that, for every $ T_k\in T $ and every $ i\in T_k $, $\mu(T_k)=\w(T_k) $ and $ \mu(i)\in B_i\big(O\backslash \bigcup_{\ell=1}^{k-1}\w(T_\ell) \big) $, and for any allocation $ \sigma $ such that $ I_{\sigma>\mu}\neq \emptyset $, there exists $ T_k\in T $ such that $ T_k \cap I_{\mu>\sigma}\neq \emptyset$ and $ k<k' $ for every $ T_{k'}\in T $ such that $ T_{k'} \cap I_{\sigma>\mu}\neq \emptyset$. We prove that $ \mu $ belongs to the exclusion core.
		
		Suppose that $ \mu $ is exclusion blocked by a coalition $ C $ via another $ \sigma $. Without loss of generality, let $ C= I_{\sigma>\mu}$. Let $ k' $ be the smallest $ \ell $ such that $ T_\ell\cap I_{\sigma>\mu}\neq \emptyset $. 		
		By the above assumption, there exists $ T_k\in T $ such that $ T_k \cap I_{\mu>\sigma}\neq \emptyset$ and $ k<k' $. Since $ k<k' $, $ T_k\cap C=\emptyset $. Since $ \mu(T_k)=\w(T_k) $, the objects in $ \w(T_k) $ are not controlled by agents outside $ T_k $. So, $ \w(T_k)\cap \Omega(C|\w,\mu)=\emptyset $. This means that there exists $ j\in T_k \cap I_{\mu>\sigma} $ but $ \mu(j)\notin  \Omega(C|\w,\mu)$, which is a contradiction. So, $ \mu $ is not exclusion blocked. 
	\end{proof}

\begin{proof}[\textbf{Proof of Proposition \ref{prop:exclusioncore:multiplecopies}}]
	In the common indifferences setting, let $ \mu $ be an outcome of TTC.  We prove that $ \mu $ belongs to the exclusion core.
	
	Let $ T=(T_1,T_2,\ldots,T_t) $ denote the order of cycles removed in the TTC procedure that produces $ \mu $, where each $ T_k\in T $ is the set of agents involved in a cycle. If multiple cycles are removed in the same step of TTC, their relative ranking can be arbitrary in  $ T $. It is obvious that, for each $ T_k\in T $ and each $ i\in T_k $, $\mu(T_k)=\w(T_k) $ and $ \mu(i)\in B_i\big(O\backslash \bigcup_{\ell=1}^{k-1}\w(T_\ell) \big) $. 
	
	Consider any allocation $ \sigma $ such that $ I_{\sigma>\mu}\neq \emptyset $. Let $ k' $ be the smallest $ \ell $ such that $ T_\ell\cap I_{\sigma>\mu}\neq \emptyset $. Consider any $ i\in  T_{k'}\cap I_{\sigma>\mu}$. Let $ x $ be the object type such that $ \sigma(i)\in O_x $. Since $ \sigma(i)\succ_i \mu(i) $ and $ \mu(i)\in B_i(O\backslash \bigcup_{\ell=1}^{k'-1}\w(T_\ell)) $, all copies of $ x $ must be assigned in the TTC procedure before $ i $ is involved in a cycle. Let $ J $ be the set of agents who receive the copies of $ x $ under $ \mu $. Since all agents in $ J $ are involved in cycles before $ i$, $ J\cap I_{\sigma>\mu}=\emptyset $. Under $ \sigma $, because $ i $ receives a copy of $ x $, it is impossible for all agents in $ J $ to receive the copies of $ x $. So, there must exist an agent $ j\in J $ who receives a copy of an object type different than $ x $ under $ \sigma $. Since agents have strict preferences over object types and $ J\cap I_{\sigma>\mu}=\emptyset $, it must be that $ j \in I_{\mu>\sigma}$. So, $ I_{\mu>\sigma}\neq 
	\emptyset $, implying that $ \mu $ is Pareto efficient. Let $ j\in T_k $. So, $ T_k \cap I_{\mu>\sigma}\neq \emptyset$ and $ k<k' $. 
	By the ``if'' result of Lemma \ref{lemma:exclusion:core:nonempty}, $ \mu $ belongs to the exclusion core.	
	
	The proof of the ``only if'' result of Lemma \ref{lemma:exclusion:core:nonempty} has shown that all elements of the exclusion core are TTC outcomes. So, the exclusion core equals the set of TTC outcomes.	
\end{proof}
	
	\begin{proof}[Proof of Proposition \ref{prop:exclusioncore}]
		
	\textbf{(1)} We first prove that the rectified strong core is a superset of the exclusion core. Given a market, let $ \mu $ be any allocation outside the rectified strong core. Suppose that $ \mu $ is rectification blocked by a coalition $ C $ via another $ \sigma $. We then prove that $ \mu $ is exclusion blocked by $C_{\sigma>\mu}$ via some allocation $\sigma'$. 	
	
	Define $ I_1=\{i\in I\backslash C:\mu(i)\in \w(C)\} $ and $ I_2=\{i\in I\backslash C:\w(i)\in \mu(C)\} $. The two sets may overlap. It is evident that $ |I_1|=|I_2| $. Let $ \sigma' $ be an allocation specifying a bijection from $I_1$ to $\w(I_2)$ such that, $ \forall i\in I\backslash (I_1 \cup C) $, $ \sigma'(i)=\mu(i) $; $ \forall i\in C $, $ \sigma'(i)=\sigma(i) $; and $ \forall i\in I_1 $, $ \sigma'(i)\in \w(I_2) $. Then, for any  $ j\in I\backslash C $ such that $ \mu(j)\succ_j \sigma'(j) $, it must be that $ j\in I_1 $ and therefore $ \mu(j)\in \w(C)$. We need to prove that $\mu(j)\in  \Omega(C_{\sigma> \mu}|\w,\mu)$. Because $C_{\sigma'\sim \mu}=C_{\sigma\sim \mu}$, for every $ i\in C_{\sigma'\sim \mu}$, it holds that $ \mathcal{I}_i(\mu(i))\subseteq \w(C) $.

	Since $ \mu(j)\in \w(C)$, either $\mu(j)\in  \w(C_{\sigma> \mu})$ or $\mu(j)\in  \w(C_{\sigma\sim \mu})$. In the former case, we are done. In the latter case, $ \mu(j)=\w(i_1) $ for some $ i_1\in C_{\sigma\sim\mu} $. Since $ \mu(i_1)\in \w(C) $, either $ \mu(i_1)\in \w(C_{\sigma>\mu}) $ or $ \mu(i_1)\in \w(C_{\sigma\sim\mu}) $. In the former case, we obtain $ \mu(j)\in \Omega(C_{\sigma>\mu}|\w,\mu) $. In the latter case, $ \mu(i_1)=\w(i_2) $ for some $ i_2\in C_{\sigma\sim\mu} $. Repeating the argument, we will find a chain $ j \rightarrow i_1 \rightarrow i_2 \rightarrow \cdots \rightarrow i_K $ in which every $ i_k $ belongs to $ C $ and every agent obtains his pointee's endowment. It is impossible that every $ i_k $ belongs to $ C_{\sigma\sim\mu} $, since otherwise they would form a cycle, contradicting that $ j$ does not belong to $C$ yet obtains $ i $'s endowment. Therefore, at least one agent in the chain belongs to $ C_{\sigma>\mu} $, which means that $ \mu(j)\in \Omega(C_{\sigma>\mu}|\w,\mu) $.

	 \textbf{(2)} We then prove that the exclusion core is a superset of the strong core. This obviously holds when the strong core is empty. When the strong core is nonempty, let $ \mu $ be any allocation within it. By Lemma \ref{lemma:strong:core:nonempty}, there exists a TTS $ T^*=(T_1,T_2,\ldots,T_{t^*}) $ and for every $ T_k\in T^* $ and every $ i\in T_k$, $ \mu(T_k)=\w(T_k) $ and $ \mu(i)\in B_i(O\backslash \bigcup_{\ell=1}^{k-1}\w(T_\ell)) $. For any allocation $ \sigma $ such that $ I_{\sigma>\mu}\neq \emptyset $, let $ k' $ be the smallest $ \ell $ such that $ T_\ell\cap I_{\sigma>\mu}\neq \emptyset $. For every $ i\in  T_{k'}\cap I_{\sigma>\mu}$, since $ \sigma(i)\succ_i \mu(i) $ and $ \mu(i)\in B_i(O\backslash \bigcup_{\ell=1}^{k'-1}\w(T_\ell)) $, it must be that	 
	  $ \sigma(i)\in \bigcup_{\ell=1}^{k'-1}\w(T_\ell) $. So, there must exist $ j\in  \bigcup_{\ell=1}^{k'-1}T_\ell $ such that $ \sigma(j)\notin \bigcup_{\ell=1}^{k'-1}\w(T_\ell) $. Let $ j\in T_k $. So $ k<k' $. Since $ T^* $ is a TTS, all objects that $ j $ views as no worse than $ \mu(j) $ are owned by $ \bigcup_{\ell=1}^{k}T_\ell $. So, $ \mu(j)\succ_j\sigma(j) $. That is, $ j\in T_k \cap I_{\mu>\sigma} $. Then, by the ``if'' result in Lemma \ref{lemma:exclusion:core:nonempty}, $ \mu $ belongs to the exclusion core.

		When the strong core is nonempty, Theorem \ref{theorem:rectifiedcore} has shown that the rectified strong core coincides with the strong core, so the three cores coincide.
	\end{proof}

\begin{proof}[\textbf{Proof of Proposition \ref{prop:rectifiedcore:multiplecopies}}]
	Proposition \ref{prop:exclusioncore:multiplecopies} shows that in the common indifferences setting, the exclusion core equals the set of TTC outcomes.	Proposition \ref{prop:exclusioncore} shows that the exclusion core is always a subset of the rectified strong core. So, when agents have common indifferences, every TTC outcome belongs to the rectified strong core.
\end{proof}

\section{Additional examples}\label{appendix:examples}

\begin{example}[Intermediate case for justifying condition (3)]\label{example:othercases}
		Consider the following market with five agents.
		
		\begin{table}[!h]
    \raggedright
    \begin{subtable}[t]{.3\linewidth}
        \begin{tabular}[t]{cccccc}
            & $ 1 $ & $ 2 $ & $ 3 $ & $ 4 $ & $ 5 $  \\ \hline
            $ \w $: & $ a $ & $ b $ & $ c $ & $ d $ & $ e $ \\ \hline
            $ \mu $: & $ b $ & $ c $ & $ d $ & $ e $ & $ a $ \\
            $ \sigma $: & $ b $ & $ a $ & $ d $ & $ c $ & $ e $\\
        \end{tabular}
    \end{subtable}%
    \begin{subtable}[t]{.3\linewidth}
        \begin{tabular}[t]{ccccc}
            $ \succsim_1 $ & $ \succsim_2 $ & $ \succsim_3 $ & $ \succsim_4 $ & $ \succsim_5 $ \\ \hline
            $ b $ & $ a,c $ & $ b,d $ & $ c $ & $ a $\\
            $ a $ & $ b $ & $ c $ & $ e $ & $ \vdots $ \\
            $ \vdots $ & $ \vdots $ & $ \vdots $ & $ \vdots $
        \end{tabular}
    \end{subtable}%
    \begin{subtable}[t]{.3\linewidth}
        \centering
				\small
        \begin{tikzpicture}[semithick,bend angle=25,xscale=0.6,yscale=0.6,baseline=(current bounding box.north)]
            \node at (5,-1.5) {$\mu$};
            \node (i1) at (3,0) [agent] {$1$};
            \node (i2) at (5,0) [agent] {$2$};
            \node (i3) at (7,0) [agent] {$3$};
            \node (i4) at (9,0) [agent] {$4$};
            \node (i5) at (1,0) [agent] {$5$};
            \draw[-latex] (i1) to (i2);
            \draw[-latex] (i2) to (i3);
            \draw[-latex] (i3) to (i4);
            \draw[-latex] (i4) [bend left] to (i5);
            \draw[-latex] (i5) to (i1);
        \end{tikzpicture}
    \end{subtable}
\end{table}

Both the strong core and the rectified strong core equal $\{\sigma\}$. Under $\sigma$, each of the four agents in $\{ 1,2,3,4\} $  receives a most preferred object.

The allocation $\mu$ is rectification blocked by $C=\{ 1,2,3,4\} $ via $\sigma$, with $C_{\sigma\sim\mu}=\{1,2,3\}$ and $C_{\sigma>\mu}=\{4\}$. The unaffected group $C_{\sigma\sim\mu}$ cannot reallocate its endowments among its members to maintain their welfare under $\mu$. Among $C_{\sigma\sim\mu}$, either the subcoalition $\{1,2\}$ or the subcoalition $\{2,3\}$ can be self-sufficient, but the two groups cannot be self-sufficient simultaneously. 

If $\{1,2\}$ can be self-sufficient, then under $\mu$, if $4$ were to reclaim $d$, then $3$ would be unable to maintain his welfare, since $\{1,2\}$ would allocate their endowments among themselves. So, $3$ can be viewed as compelled to join the coalition, and the participation of $\{1,2\}$ can be justified by the neutrality argument. 

If $\{2,3\}$ can be self-sufficient, then under $\mu$, if $4$ were to reclaim $d$, then $1$ would be unable to maintain his welfare, since $\{2,3\}$ would allocate their endowments among themselves. So, $1$ can be viewed as  compelled to join the coalition, and the participation of $\{2,3\}$ can be justified by the neutrality argument.
\end{example}

\begin{example}[$\emptyset \neq $ Strong core $\subsetneq$ Bargaining set]\label{example:bargainingset}
	
	Consider a market with five agents.
	
	\begin{table}[!h]
		\centering
		\begin{subtable}[t]{.3\linewidth}
			\begin{tabular}[t]{cccccc}
				& $ 1 $ & $ 2 $ & $ 3 $ & $ 4 $ & $ 5 $  \\ \hline
				$ \w $: & $ a $ & $ b $ & $ c $ & $ d $ & $ e $ \\ \hline
				$ \mu $: & $ e $ & $ b $ & $ a $ & $ c $ & $ d $\\
				$ \delta $: & $ b $ & $ a $ & $ c $ & $ e $ & $ d $ \\
				& 
			\end{tabular}
		\end{subtable}
		\begin{subtable}[t]{.3\linewidth}
			\begin{tabular}[t]{ccccc}
				$ \succsim_1 $	& $ \succsim_2 $ & $ \succsim_3 $ & $ \succsim_4 $ & $ \succsim_5 $ \\ \hline
				$ b $ & $ a,b $ & $ a $ & $ c,e $ & $ d $\\
				$ e $ & $ \vdots $ & $ c$ & $ \vdots $ & $ \vdots $ \\
				$ \vdots $& & $ \vdots $ & 
			\end{tabular}
		\end{subtable}
	\end{table}
	
	The strong core equals $\{\delta\}$. We show that $\mu$, which is not in the strong core, belongs to the bargaining set. In $\mu$, $2$ receives his endowment, and the remaining agents exchange endowments along the cycle $1\rightarrow 5 \rightarrow 4 \rightarrow 3 \rightarrow 1$. The only coalition that can weakly block $\mu$ is $\{1,2\}$, via an allocation $\sigma$ in which they exchange endowments. By the selection rule of \cite{yilmaz2022stability}, all other agents receive their own endowments under $\sigma$. Then, $\sigma$ is weakly blocked by $\{2,4,5\}$ via an allocation $\mu'$ in which $2$ receives his endowment, and $4$ and $5$ exchange their endowments. The counter-blocking coalition $\{2,4,5\}$ overlaps with $\{1,2\}$, and every member of $\{2,4,5\}$ receives under $\mu'$ an object indifferent to his assignment under $\mu$. So, $\mu$ belongs to the bargaining set.   
\end{example}

\begin{example}[Rectified strong core $ \nsubseteq $ bargaining set]\label{example:bargainingset:2}
	Consider the following market.
	
	\begin{table}[H]
		\centering
		\begin{subtable}[b]{.3\linewidth}
			\begin{tabular}[b]{ccccc}
				& $ 1 $ & $ 2 $ & $ 3 $ & $ 4 $  \\ \hline
				$ \w $: & $ a $ & $ b $ & $ c $ & $ d $ \\ \hline
				$ \mu $: & $ b $ & $ a $ & $ d $ & $ c $\\
				$ \sigma $: & $ c $ & $ a $ & $ b $ & $ d $\\
				& 
			\end{tabular}
		\end{subtable}
		\begin{subtable}[b]{.3\linewidth}
			\begin{tabular}[b]{cccc}
				$ \succsim_1 $	& $ \succsim_2 $ & $ \succsim_3 $ & $ \succsim_4 $  \\ \hline
				$ c $ & $ a,b,c,d $ & $ b $ & $ c $ \\
				$ b $ &  & $ d $ & $ d $ \\
				$ a $ &  & $ c $ & $ a $\\
				$ d $ & & $ a $ & $ b $
			\end{tabular}
		\end{subtable}
	\end{table}
	
	We show that $ \mu $ belongs to the rectified strong core but not to the bargaining set. In $\mu$, the two pairs $\{1,2\}$ and $\{3,4\}$ each exchange their endowments.
	
	To show that $ \mu $ belongs to the rectified strong core, suppose it is rectification blocked by a coalition $ C $ via some $ \mu' $. Since $ \mu $ is Pareto efficient, $ C\subsetneq I $. Since $ 2 $ is indifferent among all objects, $ 2\notin C $. Since $ 1 $ and $ 3 $ are the only agents who can be made strictly better off, $ C_{\mu'>\mu}\subseteq \{1,3 \}$. If $ 3\in C_{\mu'>\mu} $, then $ 3 $ must receive $ b $ in $ \mu' $, which requires $ 2 \in C$, a contradiction. So, $ 1\in C_{\mu'>\mu} $, which requires $ \mu'(1)=c $. Since $ \mu(4)=c $ and $ 4 $ most prefers $ c $, we need $ 3\in C $ and $ 4\notin C $. So, $ C=\{1,3\} $ and $ C_{\mu'\sim\mu} =\{3\}$, implying $ \mu'(3)=\mu(3)=d $. But this contradicts $ 4\notin C $.

	To show that $ \mu $ does not belong to the bargaining set, note that $ \mu $ is only weakly blocked by $ \{1,2,3\} $ via $ \sigma $, and $ \sigma $ is only weakly blocked by $ \{2,3,4\} $ via another $ \mu' $ in which the three agents exchange their endowments along the cycle $ 3\rightarrow 2 \rightarrow 4 \rightarrow 3 $. However, since  $ \mu'(3)\succ_3 \mu(3) $, the blocking by $ \{2,3,4\} $ does not satisfy the requirement that coalition members must block by claiming their original welfare under $ \mu $. So, there does not exist a counter-blocking against the blocking by $\{1,2,3\}$, which means that $\mu$ does not belong to the bargaining set.
\end{example}

\setlength{\bibsep}{0pt plus 0.3ex}
\bibliographystyle{aer}

\bibliography{reference}

@book{vonNeumann1944,
	author =	 {John von Neumann and Oskar Morgenstern},
	title =	 {Theory of Games and Economic Behavior},
	publisher =	 {Princeton University Press},
	address =	 {Princeton, NJ, USA},
	year =	 {1944}
}

@article{aslan2020competitive,
  title={Competitive equilibria in Shapley--Scarf markets with couples},
  author={Aslan, Fatma and Lain{\'e}, Jean},
  journal={Journal of Mathematical Economics},
  volume={89},
  pages={66--78},
  year={2020},
  publisher={Elsevier}
}

@article{cheng2024proper,
	title={Proper Exclusion Right, Priority and Allocation of Positions},
	author={Cheng, Yao and Yang, Zaifu and Yu, Jingsheng},
	year={2024},
	journal={working paper}
}

@article{roth2005pairwise,
	title={Pairwise kidney exchange},
	author={Roth, Alvin E and S{\"o}nmez, Tayfun and {\"U}nver, M Utku},
	journal={Journal of Economic theory},
	volume={125},
	number={2},
	pages={151--188},
	year={2005},
	publisher={Elsevier}
}

@article{bogomolnaia2004random,
	title={Random matching under dichotomous preferences},
	author={Bogomolnaia, Anna and Moulin, Herv{\'e}},
	journal={Econometrica},
	volume={72},
	number={1},
	pages={257--279},
	year={2004},
	publisher={Wiley Online Library}
}

@article{ishida2025group,
	title={Group incentive-compatible allocation of discrete resources when ownership is partitioned},
	author={Ishida, Wataru and Park, Changwoo},
	journal={Games and Economic Behavior},
	year={2025},
	publisher={Elsevier}
}

@article{sandholtz2025shapley,
	title={Shapley-Scarf Markets with Objective Indifferences},
	author={Sandholtz, Will and Tai, Andrew},
	journal={arXiv preprint arXiv:2503.18144},
	year={2025}
}

@article{afacan2024housing,
	title={Housing markets since Shapley and Scarf},
	author={Afacan, Mustafa O{\u{g}}uz and Hu, Gaoji and Li, Jiangtao},
	journal={Journal of Mathematical Economics},
	volume={111},
	pages={102967},
	year={2024},
	publisher={Elsevier}
}

@article{toda1997implementation,
	title={Implementation and characterizations of the competitive solution with indivisibility},
	author={Toda, M},
	year={1997},
	journal={mimeo}
}

@article{wako1999coalition,
	title={Coalition-proofness of the competitive allocations in an indivisible goods market},
	author={Wako, Jun},
	journal={Fields Institute Communications},
	volume={23},
	pages={277--283},
	year={1999}
}

@article{ehlers2004monotonic,
	title={Monotonic and implementable solutions in generalized matching problems},
	author={Ehlers, Lars},
	journal={Journal of Economic Theory},
	volume={114},
	number={2},
	pages={358--369},
	year={2004},
	publisher={Elsevier}
}

@article{klaus2010farsighted,
	title={Farsighted house allocation},
	author={Klaus, Bettina and Klijn, Flip and Walzl, Markus},
	journal={Journal of Mathematical Economics},
	volume={46},
	number={5},
	pages={817--824},
	year={2010},
	publisher={Elsevier}
}

@article{kawasaki2010farsighted,
	title={Farsighted stability of the competitive allocations in an exchange economy with indivisible goods},
	author={Kawasaki, Ryo},
	journal={Mathematical Social Sciences},
	volume={59},
	number={1},
	pages={46--52},
	year={2010},
	publisher={Elsevier}
}

@article{aumann1964bargaining,
	title={The bargaining set for cooperative games},
	author={Aumann, Robert J and Maschler, Michael},
	journal={Advances in game theory},
	volume={52},
	number={1},
	pages={443--476},
	year={1964},
	publisher={Princeton University Press Princeton}
}

@article{biro2023shapley,
	title={Shapley--Scarf Housing Markets: Respecting Improvement, Integer Programming, and Kidney Exchange},
	author={Bir{\'o}, P{\'e}ter and Klijn, Flip and Klimentova, Xenia and Viana, Ana},
	journal={Mathematics of Operations Research},
	year={2023},
	publisher={INFORMS}
}

@article{demuynck2019myopic,
	title={The myopic stable set for social environments},
	author={Demuynck, Thomas and Herings, P Jean-Jacques and Saulle, Riccardo D and Seel, Christian},
	journal={Econometrica},
	volume={87},
	number={1},
	pages={111--138},
	year={2019},
	publisher={Wiley Online Library}
}

@article{wako2007nonexistence,
	title={On the nonexistence of vNM stable sets in an exchange economy with indivisible goods},
	author={Wako, Jun and Matsumoto, Kiiko and Irisawa, Toshiharu},
	year={2007},
	journal={mimeo}
}

@article{yilmaz2022stability,
	title={Stability of an allocation of objects},
	author={Y{\i}lmaz, Murat and Y{\i}lmaz, {\"O}zg{\"u}r},
	journal={Review of Economic Design},
	volume={26},
	number={4},
	pages={561--580},
	year={2022},
	publisher={Springer}
}

@article{erdil2017two,
	title={Two-sided matching with indifferences},
	author={Erdil, Aytek and Ergin, Haluk},
	journal={Journal of Economic Theory},
	volume={171},
	pages={268--292},
	year={2017},
	publisher={Elsevier}
}

@article{bogomolnaia2005strategy,
	title={Strategy-proof assignment on the full preference domain},
	author={Bogomolnaia, Anna and Deb, Rajat and Ehlers, Lars},
	journal={Journal of Economic Theory},
	volume={123},
	number={2},
	pages={161--186},
	year={2005},
	publisher={Elsevier}
}

@article{mumcu2007core,
	title={The core of a housing market with externalities},
	author={Mumcu, Ayse and Saglam, Ismail},
	journal={Economics Bulletin},
	volume={3},
	number={55},
	pages={1--5},
	year={2007}
}

@article{graziano2020shapley,
	title={Shapley and Scarf housing markets with consumption externalities},
	author={Graziano, Maria Gabriella and Meo, Claudia and Yannelis, Nicholas C},
	journal={Journal of Public Economic Theory},
	volume={22},
	number={5},
	pages={1481--1514},
	year={2020},
	publisher={Wiley Online Library}
}

@article{dougan2011core,
	title={The Core of Shapley--Scarf markets with couples},
	author={Do{\u{g}}an, Onur and Laffond, Gilbert and Lain{\'e}, Jean},
	journal={Journal of Mathematical Economics},
	volume={47},
	number={1},
	pages={60--67},
	year={2011},
	publisher={Elsevier}
}

@article{klaus2023core,
	title={The core for housing markets with limited externalities},
	author={Klaus, Bettina and Meo, Claudia},
	journal={Economic Theory},
	pages={1--33},
	year={2023},
	publisher={Springer}
}

@article{hong2022core,
	title={Core and top trading cycles in a market with indivisible goods and externalities},
	author={Hong, Miho and Park, Jaeok},
	journal={Journal of Mathematical Economics},
	volume={100},
	pages={102627},
	year={2022},
	publisher={Elsevier}
}

@inproceedings{plaxton2013simple,
	title={A simple family of top trading cycles mechanisms for housing markets with indifferences},
	author={Plaxton, C Gregory},
	booktitle={Proceedings of the 24th international conference on game theory},
	pages={1--23},
	year={2013},
	organization={Citeseer}
}

@article{wako1984note,
	title={A note on the strong core of a market with indivisible goods},
	author={Wako, Jun},
	journal={Journal of Mathematical Economics},
	volume={13},
	number={2},
	pages={189--194},
	year={1984},
	publisher={Elsevier}
}

@article{wako1991some,
	title={Some properties of weak domination in an exchange market with indivisible goods},
	author={Wako, Jun},
	journal={The Economic Studies Quarterly},
	volume={42},
	number={4},
	pages={303--314},
	year={1991},
	publisher={JAPANESE ECONOMIC ASSOCIATION}
}

@inproceedings{aziz2012housing,
	title={Housing markets with indifferences: A tale of two mechanisms},
	author={Aziz, Haris and De Keijzer, Bart},
	booktitle={Proceedings of the AAAI Conference on Artificial Intelligence},
	volume={26},
	number={1},
	year={2012}
}

@article{ahmad2021group,
	title={Group incentive compatibility in the housing market problem with weak preferences},
	author={Ahmad, Ghufran},
	journal={Games and Economic Behavior},
	volume={126},
	pages={136--162},
	year={2021},
	publisher={Elsevier}
}

@inproceedings{saban2013house,
	title={House allocation with indifferences: a generalization and a unified view},
	author={Saban, Daniela and Sethuraman, Jay},
	booktitle={Proceedings of the fourteenth ACM conference on Electronic Commerce},
	pages={803--820},
	year={2013}
}

@article{zhangrefinedexclusioncore,
	title={Consistency and the exclusion core in object allocation with co-ownership},
	author={Zhang, Jun},
	journal={working paper},
	year={2026}
}

@article{quint2004houseswapping,
	title={On houseswapping, the strict core, segmentation, and linear programming},
	author={Quint, Thomas and Wako, Jun},
	journal={Mathematics of Operations Research},
	volume={29},
	number={4},
	pages={861--877},
	year={2004},
	publisher={INFORMS}
}

@article{jaramillo2012difference,
	title={The difference indifference makes in strategy-proof allocation of objects},
	author={Jaramillo, Paula and Manjunath, Vikram},
	journal={Journal of Economic Theory},
	volume={147},
	number={5},
	pages={1913--1946},
	year={2012},
	publisher={Elsevier}
}

@article{alcalde2011exchange,
	title={Exchange of indivisible goods and indifferences: The top trading absorbing sets mechanisms},
	author={Alcalde-Unzu, Jorge and Molis, Elena},
	journal={Games and Economic Behavior},
	volume={73},
	number={1},
	pages={1--16},
	year={2011},
	publisher={Elsevier}
}

@article{gale1962college,
	title={College admissions and the stability of marriage},
	author={Gale, David and Shapley, Lloyd S},
	journal={American mathematical monthly},
	pages={9--15},
	year={1962},
	publisher={JSTOR}
}

@article{balbuzanov2019property,
	title={The property rights theory of production networks},
	author={Balbuzanov, Ivan and Kotowski, Maciej H},
	journal={Theoretical Economics},
	volume={19},
	number={4},
	pages={1619--1658},
	year={2024},
	publisher={Wiley Online Library}
}

@article{balbuzanov2019endowments,
	title={Endowments, exclusion, and exchange},
	author={Balbuzanov, Ivan and Kotowski, Maciej H},
	journal={Econometrica},
	volume={87},
	number={5},
	pages={1663--1692},
	year={2019},
	publisher={Wiley Online Library}
}

@article{roth1977weak,
	title={Weak versus strong domination in a market with indivisible goods},
	author={Roth, Alvin E and Postlewaite, Andrew},
	journal={Journal of Mathematical Economics},
	volume={4},
	number={2},
	pages={131--137},
	year={1977},
	publisher={Elsevier}
}

@article{shapley1974cores,
	title={On cores and indivisibility},
	author={Shapley, Lloyd and Scarf, Herbert},
	journal={Journal of Mathematical Economics},
	volume={1},
	number={1},
	pages={23--37},
	year={1974},
	publisher={Elsevier}
}

\end{document}